\documentclass[a4paper,12pt]{article}
\usepackage{amsfonts, amsmath,amssymb,bm,enumerate, graphicx,color}
\usepackage{subfigure}
\usepackage{epstopdf}
\usepackage[dvipsnames]{xcolor}
\usepackage{float}
\usepackage{graphicx}			
\usepackage{mathrsfs}
\usepackage{amsfonts}
\usepackage{amsmath,amssymb,mathrsfs,bm,latexsym,graphicx,color}		
\usepackage{amssymb}
\usepackage[dvipsnames]{xcolor}
\usepackage[normalem]{ulem}
\usepackage[colorlinks=true,urlcolor=blue,citecolor=blue,linkcolor=purple,bookmarks=true]{hyperref}
\usepackage{tikz}
\usepackage{multicol}
\usepackage{multirow}
\usepackage{enumitem}

\usepackage[round]{natbib}
\usepackage[T1]{fontenc}
\usepackage[utf8]{inputenc}

\allowdisplaybreaks

\newcounter{quest}

\usepackage{footnote}
\usepackage[affil-it]{authblk}
\usepackage[english]{babel}
\usepackage{blindtext}

\newtheorem{theorem}{Theorem}
\newtheorem{proposition}[theorem]{Proposition}
\newtheorem{lemma}[theorem]{Lemma}

\newtheorem{example}[theorem]{Example}

\numberwithin{theorem}{section}
\allowdisplaybreaks

\newcommand{\cl}[1]{\mathcal{#1}}

\newcommand{\bR}{\mathbb{R}}

\newcommand{\bQ}{\mathbb{Q}}

\begin{document}

\title{\bf Risk-minimizing reinsurance with adverse selection}


\author{{\small
Ka Chun Cheung\textsuperscript{a},
Sheung Chi Phillip Yam\textsuperscript{b},
Fei Lung Yuen\textsuperscript{c}, and
Yiying Zhang\textsuperscript{d}%
\thanks{\scriptsize
\textsuperscript{a}Department of Statistics and Actuarial Science, The University of Hong Kong, Hong Kong (\texttt{kccg@hku.hk});
\textsuperscript{b}Department of Statistics, The Chinese University of Hong Kong, Hong Kong (\texttt{scpyam@sta.cuhk.edu.hk});
\textsuperscript{c}Department of Statistics and Actuarial Science, University of Kent, United Kingdom (\texttt{k.yuen@kent.ac.uk});
\textsuperscript{d}Department of Mathematics, Southern University of Science and Technology, Shenzhen, China (\texttt{zhangyy3@sustech.edu.cn}).}}}

\maketitle

\begin{abstract}
This paper provides a comprehensive characterization of optimal reinsurance mechanisms under adverse selection when a monopolistic reinsurer faces a continuum of insurers with multi-dimensional private information on loss exposures and Value-at-Risk (V$@$R) preferences. Moving beyond standard single-dimensional or discrete-type models, we resolve an infinite-dimensional, non-convex screening problem by introducing a novel pre-reinsurance V$@$R reparameterization and an indirect utility representation. This framework converts global incentive constraints into tractable, typewise conditions. Using a Breeden–Litzenberger representation, we show that optimal indirect utility consistently adopts a hinge form across stop-loss, quota-share, and change-loss contract classes, yielding an endogenous market-exclusion threshold. Among participating types, stop-loss contracts screen via type-dependent deductibles, whereas quota-share contracts induce complete pooling at full coverage. Moreover, under mild conditions, expanding the contract space to change-loss policies yields no extra profit beyond optimal stop-loss designs. Numerical examples further illustrate the fundamental economic trade-off between extracting information rents and excluding low-exposure types. Our approach provides deeper insight into the mechanisms of discrimination and exclusion in reinsurance markets characterized by asymmetric information, enriching theoretical and practical perspectives on contract design in the presence of adverse selection.

\medskip

\end{abstract}

\section{Introduction}
Reinsurance enables insurers to transfer part of their underwriting risk to reinsurers. The modern theory of optimal reinsurance dates back to the seminal work of \cite{borch1960attempt}, which established the optimality of stop-loss treaties under the expected-value premium principle. Since then, a large literature has examined optimal insurance and reinsurance design under alternative preferences, premium principles, and economics background; see, for example, \cite{arrow1963uncertainty,borch1969optimal,young1999optimal,cai2007optimal,bernard2015optimal,zhuang2016marginal,asimit2017robust,xu2019optimal,chi2020optimal,birghila2023optimal}.

Most of the existing literature on optimal (re)insurance design assumes that the relevant characteristics of the contracting parties are commonly observed, including loss distributions, risk preferences, and premium principles. However, in practice, (re)insurance decisions may involve multidimensional private information, including both privately known loss risk and heterogeneity in preferences and other demand-relevant characteristics \citep[see, e.g.,][]{finkelstein2004adverse,finkelstein2006multiple,cohen2010testing}. Such information asymmetries can lead to adverse selection before contracting and moral hazard after contracting, and may call for signaling or screening mechanisms. These issues have long been central to contract theory and principal--agent models \citep[cf.][]{talmor1981asymmetric,zhang2012dynamic,dosis2018signalling,zhang2022signaling}.

In this paper, we deal with adverse selection on the insurer's side of the reinsurance market. In particular, an insurer may have private information about its loss exposure, risk preferences, or other characteristics that matter for the value of reinsurance but cannot be directly observed by the reinsurer. In this case, a reinsurer with market power must then offer contracts that induce insurers to reveal this information through their choices. Under the framework of  adverse selection through privately known loss probabilities,  \cite{rothschild1976equilibrium} study competitive insurance policies, while \cite{stiglitz1977monopoly} considers nonlinear pricing by a monopolistic insurer, both in fixed-loss settings with a common Bernoulli utility function. \cite{chade2012optimal} extend the monopoly problem to a continuum of privately informed risk types, allowing for wealth effects and arbitrary type distributions. Beyond one-dimensional risk heterogeneity, \cite{landsberger1999general} allow two types to differ jointly in their loss distributions and Bernoulli utility functions and show that certainty equivalents can be used to order the types. \cite{crocker2011multidimensional} show, from a different perspective, how multiple coverages, perils, deductibles, and policy limits can be used to screen along several dimensions of private information. More recently, \cite{farinha2023risk} study competitive insurance with a continuum of privately informed risk and preference types, with particular attention to risk classification and welfare. 

The above mentioned work mainly considers the optimal design of (re)insurance menu with adverse selection under the framework of expected utility. From the perspective of risk management, there also exists a small branch of relevant research work. For example, \cite{cheung2019reinsurance} consider insurers that evaluate risk using V$@$R and show that combining quota-share and stop-loss features can mitigate adverse selection. The optimal menu depends on both the composition of the market and the degree of risk heterogeneity. \cite{cheung2020concave} extend this framework to menus of stop-loss contracts under concave distortions and characterize the resulting optimal deductibles. Without restricting indemnities to a parametric family, \cite{liang2022revisiting} derive optimal contract menus when insurers' distortion risk measures overstate tail risk. \cite{Boonenzhang2021} consider a related problem in which the reinsurer is uncertain about the insurer's distortion risk measure and holds probabilistic beliefs over its possible specifications. These papers provide natural benchmarks for asymmetric-information reinsurance with a focus on two types of private policyholders, but the continuum-type case remains much less understood. Along this line, \cite{gershkov2023optimal} study heterogeneous loss distributions under a common Yaari's dual utility for a continuum of agents, whereas \cite{ghossoub2025optimal} consider privately known risk attitudes under a common loss distribution. More recently, \cite{han2026optimal} analyze uncertainty about loss distributions and risk attitudes separately in a mean--variance Stackelberg framework. 

However, less attention has been paid to reinsurance markets in which insurers privately know both dimensions of their loss distributions and preference types \citep[cf.][]{rochet1998ironing,olivella2014multidimensional,carroll2017robustness}. To address this gap, we study a monopolistic reinsurer that offers a menu of type-dependent, loss-contingent indemnity contracts to a continuum of insurers who privately know both their loss exposures and V$@$R probability levels. The reinsurer evaluates ceded losses through a potentially nonlinear risk-adjusted valuation functional and chooses the menu to maximize profit. We analyze three common contract classes: stop-loss, quota-share, and change-loss policies. Across all three classes, the reinsurer sets an endogenous cutoff in pre-reinsurance V$@$R, excludes insurers below the cutoff, and serves those above it. Although adverse-selection models with participation constraints often generate market exclusion \citep[cf.][]{jullien2000participation,guerrieri2010adverse}, our model explicitly links the cutoff to V$@$R-based preferences and reinsurance costs. Among participating insurers, stop-loss contracts generate type-dependent deductibles, whereas quota-share contracts pool insurers at full reinsurance. Additionally, under the stated structural condition, allowing the more flexible change-loss class leaves the optimal stop-loss allocation unchanged and does not increase the reinsurer's profit.

The mathematical challenge arises from the infinite-dimensional and generally non-convex nature of the screening problem, together with the presence of global incentive constraints. To overcome this difficulty, we exploit the structure induced by V$@$R: across all three contract classes, an insurer's valuation depends on its private characteristics $(\alpha,k)$ only through its V$@$R, i.e., $a=\text{V$@$R}_{\alpha}(X_k)$, which is  called the \textit{pre-reinsurance} V$@$R, standing for the insurer's gross loss before reinsurance. We therefore use $(a,k)$ as the type variable and represent the premium menu through a univariate indirect utility function $v(a)$.\footnote{Interested readers can refer to \cite{carlier2007optimal,horst2008risk,cvitanic2022optimal} for more details on technical usage of indirect utility functions in some relevant optimization problems arising from insurance and finance.}  Incentive compatibility requires $v$ to be increasing, convex, and one-Lipschitz, which reduces the global constraints to typewise conditions. This reparameterization preserves cost heterogeneity: insurers with the same $a$ value contracts identically, while the reinsurer's cost $H[I(X_k)]$ may still vary with $k$. We exploit this separation by fixing $v$ and optimizing indemnities pointwise. Under stop-loss reinsurance, incentive compatibility yields a hinge-shaped $v$; for quota-share and change-loss contracts, the Breeden--Litzenberger representation expresses admissible $v$ as mixtures of hinge functions $\phi_t(a)=(a-t)_+$.

The remainder of the paper is organized as follows. Section \ref{sec:model} formulates the optimal reinsurance menu design problem under adverse selection. Section \ref{sec:failure} illustrates the failure of first-best contracts under asymmetric information, while Section \ref{sec:indirect} reformulates the problem using indirect utility. Sections \ref{sec:stoploss}, \ref{sec:quotashare}, and \ref{sec:changeloss} characterize the optimal menus under stop-loss, quota-share, and change-loss reinsurance, respectively, together with numerical illustrations. Section \ref{sec:con} concludes the paper.

\section{Model formulation}\label{sec:model}
Throughout, we denote $\mathbb{R}=(-\infty,\infty)$, $\mathbb{R}_+=[0,\infty)$ and $\overline{\mathbb{R}}_+=\mathbb{R}_+\cup\{\infty\}$. Let  $(K, \cl{K})$ be a measurable space. We assume that $K$ is a Polish space, and $\cl{K}$ is the Borel $\sigma$-algebra. In this paper, we assume that types of agents are represented by the pairs $(\alpha,k)\in(0,1)\times K$, and the collection of all possible types of agents in the market is denoted by $\tilde{\cl{T}}\subset (0,1)\times K$. We assume that
 $\tilde{\cl{T}}$ is measurable with respect to the product $\sigma$-algebra $\cl{B}(0,1)\otimes\cl{K}$, and
\[\{k\in K: \exists \alpha\in(0,1)\text{ such that }(\alpha,k)\in\tilde{\cl{T}}\}=K,\]
so that every element in $K$ corresponds to at least one agent. Each agent can be identified with his type, and for agent $(\alpha,k)$, we suppose that he is exposed to the positive loss variable $X_k$ whose distribution function and survival function are $F_k$ and $\overline{F}_k$ respectively, and is a $\text{V$@$R}_\alpha (= \overline{F}^{-1}(\alpha))$ minimizer. For simplicity, it is assumed that each $F_k$ is   strictly increasing and continuous on its support, with possibly a jump (point mass) at 0. To avoid triviality, we further assume that  $\alpha< 1-F_k(0)$ for any $(\alpha,k)\in\tilde{\cl{T}}$. These assumptions ensure that the map $\iota: \tilde{\cl{T}}\to \bR_+\times K$ defined by
\[ \iota (\alpha,k) := (\text{V$@$R}_\alpha (X_k) ,k), \]
is injective; we further assume that the map $\iota$ is measurable.


Following standard literature on adverse selection, it is assumed that the type of each agent is his own private information, hidden from the reinsurer. Nevertheless, the reinsurer does know the distribution of types across the concerned population. Such distribution is represented by the probability measure
$\tilde{\mathbb{Q}}$ on $\tilde{\cl{T}}$. The objective of the reinsurer is to design a menu of policies for agents to freely choose from, so that the expected profit is maximized.

According to the revelation principle \citep[cf.][]{macho2001introduction,laffont2009theory},
the reinsurer may restrict the menu of
contracts to those that provide each type of agent with the
incentive to truthfully reveal his characteristic. In other words, one only needs to consider menu of policies of the form
\[
M=\{(I_{(\alpha,k)},P_{(\alpha,k)})\mid (\alpha,k)\in \tilde{\cl{T}}\},
\]
and the expected profit of the reinsurer is expressed as
\begin{equation}\label{ExProfit}
J[M]:=\int_{\tilde{\cl{T}}}(P_{(\alpha,k)} - H[I_{(\alpha,k)}(X_k)])\,\tilde{\mathbb{Q}}(d\alpha\times dk),
\end{equation}
Here, the functional $H:\cl{X}\to\bR$ represents the reinsurer's risk-adjusted valuation of an indemnity payment, incorporating the expected claim cost together with any risk or capital loading. Thus, $H[I_{(\alpha,k)}(X_k)]$ is the economic cost assigned by the reinsurer to the indemnity paid to type $(\alpha,k)$. Its domain $\cl{X}$ contains all loss and indemnity risks considered below. We assume that $H$ is monotone, convex, comonotonic additive, and positive: $H(X)\geq0$ for every nonnegative $X$, with equality only when $X=0$ almost surely, and that $H(1)=1+\theta>1$. One notable example is given by
\begin{equation}\label{H:Example}
H[Y] = (1+\theta)\int_0^\infty h(\overline{F}_Y(y))\,dy,
\end{equation}
where $h$ is a concave distortion function satisfying $h(0)=0$ and $h(1)=1$.\footnote{Set $\theta=0$, this is well-known as the Wang's premium principle; see \cite{wang1996premium,young1999optimal}.}

With these notations, the integrand in (\ref{ExProfit}), $P_{(\alpha,k)}-H[I_{(\alpha,k)}(X_k)]$, is the reinsurer's profit from the policy selected by an agent of type $(\alpha,k)$. Integrating this profit with respect to $\tilde{\mathbb{Q}}$ gives the reinsurer's expected profit across the population. Equation (\ref{ExProfit}) therefore presumes that each type selects the policy designated for it. The incentive-compatibility and individual-rationality constraints introduced below ensure, respectively, truthful self-selection from the menu and voluntary participation.

As $H$ is monotone and convex, for any $\kappa\in(0,1)$ and $d_1,d_2\in\bR_+$ and for any $X\in L^1_+$,
\begin{align*}
    H[(X-(\kappa d_1+(1-\kappa)d_2))_+]&\leq H[\kappa(X-d_1)_++(1-\kappa)(X-d_2)_+]\\
    &\leq \kappa H[(X-d_1)_+]+(1-\kappa)H[(X-d_2)_+].
\end{align*}
In particular, this implies that the map
\[d\mapsto B_X(d):= -d-H[(X-d)_+]\]
is concave on $\bR_+$.
For any $k\in\cl{K}$, we denote by $\theta_k^*$ the point where $B_{X_k}$ attains its maximum value. If $B_{X_k}$ is increasing on the whole $\bR_+$, $\theta_k^*$ is simply  set as $+\infty$. By concavity, $B_{X_k}$ is increasing on $[0,\theta_k^*]$.  When the functional $H$ is given by (\ref{H:Example}), and $\theta^*_k$ can be computed explicitly by
\[\theta^*_k=(h\circ \overline{F}_k)^{-1}\left(\frac{1}{1+\theta}\right),\]
which represents some kind of distorted quantile of $X_k$.
The negative of the maximum value of $B_{X_k}$ is denoted by $\xi_k$, that is,
\[\xi_k :=-B_{X_k}(\theta^*_k)= \theta_k^*+H[(X_k-\theta_k^*)_+].\]
As we will see in subsequent sections, $\theta^*_k$ and $\xi_k$ play a key role in characterizing the optimal menus of policies.

Returning to the expression of the expected profit as given in (\ref{ExProfit}), it only makes sense if one can ensure that all agents, irrespective of their types, will self-select the particular policies that are tailor-made for them, that is, type $(\alpha,k)$ must select policy $(I_{(\alpha,k)},P_{(\alpha,k)})$ for every $(\alpha,k)\in\tilde{T}$. This can be achieved by imposing the following constraints:
\begin{itemize}
    \item [(i)] \textbf{Incentive compatibility} (IC) constraint:
\[
\text{V$@$R}_\alpha (X_k - I_{(\alpha,k)}(X_k) )+P_{(\alpha,k)}
\leq \text{V$@$R}_\alpha (X_k - I_{(\alpha',k')}(X_k) )+P_{(\alpha',k')}
\]
for any $(\alpha,k),(\alpha',k')\in \tilde{\cl{T}}$, and
\item [(ii)] \textbf{Individual rationality} (IR) constraint:
\[
\text{V$@$R}_\alpha (X_k - I_{(\alpha,k)}(X_k) )+P_{(\alpha,k)}
\leq \text{V$@$R}_\alpha (X_k)
\]
for any $(\alpha,k)\in \tilde{\cl{T}}$, which, by applying comonotonicity and increasing transformation of V$@$R, is equivalent to
\[I_{(\alpha,k)}(\text{V$@$R}_\alpha (X_k) )-P_{(\alpha,k)}\geq 0.\]
\end{itemize}
Therefore, the objective of the reinusrer is to design an optimal menu of policies that maximizes the expected profit in (\ref{ExProfit}) subject to the IC and IR constraints.

One can considerably simplify the problem by a simple change of variable. Recall that the map $\iota:\tilde{\cl{T}}\to \mathbb{R}_+\times K $ defined by
\[\iota (\alpha,k) = (\text{V$@$R}_\alpha (X_k) ,k)\]
is injective and measurable. Because of the injectivity of the map $\iota$, we can express any type $(\alpha,k)$ equivalently by $(a,k):= (\text{V$@$R}_\alpha (X_k),k)$, leading to $\iota (\alpha,k)=(a,k)$. This new representation of types has the advantages of encoding the $\alpha$-V$@$R of $X_k$ directly and simplifying the IC and IR constraints substantially.
Under this new representation, the collection of all agents' possible types becomes $\cl{T}:=\iota(\tilde{\cl{T}})\subset\bR_+\times K$.  We note that as both $(0,1)\times K$ and $\bR_+\times K$ are Polish spaces\footnote{Recall that countable product of Polish spaces is again a Polish space.}, $\cl{T}$ is measurable in $\bR_+\times K$ \citep[cf. Theorem 8.3.7 in][]{cohn2013measure}.
Under the transformation $\iota$, the distribution of types becomes
\[\mathbb{Q}(B) := \tilde{\mathbb{Q}}\circ\iota^{-1}(B), \quad B\text{ measurable in $\mathbb{R}_+\times K$},\]
which is just the image measure of $\tilde{\mathbb{Q}}$ on $\mathbb{R}_+\times K$ under the map $\iota$.
Since $\mathbb{Q}$ is supported by   $\mathcal{T}$, we may consider pairs $(a,k)\in (\mathbb{R}_+\times K)\setminus \mathcal{T}$ as the types of ``fictitious agents''. Of course, the total measure of all fictitious agents under $\mathbb{Q}$ is zero.
With this new representation of agents' types, the IC and IR constraints can now be written as
\begin{equation}\label{IC-1}
    I_{(a,k)}(a)-P_{(a,k)}\geq I_{(a',k')}(a)-P_{(a',k')}
\end{equation}
and
\begin{equation}\label{IR-1}
    I_{(a,k)}(a)-P_{(a,k)}\geq 0
\end{equation}
for any $(a,k),(a',k')\in \mathbb{R}_+\times K
$. A menu $M=\{(I_{(a,k)},P_{(a,k)})\mid (a,k)\in \mathbb{R}_+\times K\}$ satisfying the IC and IR constraints (\ref{IC-1})-(\ref{IR-1}) is called \textit{feasible}.

In this paper, we will consider the following three classes of policies:
\begin{itemize}
    \item[(i)] the class of stop-loss policies:
    \[\mathcal{I}_{1}:=\{I(x)=(x-d)_+; d\in\overline{\mathbb{R}}_+\};\]
    \item[(ii)] the class of quota share policies:
    \[\mathcal{I}_{2}:=\{I(x)=\lambda x; \lambda\in[0,1]\};\]
    \item[(iii)] the class of change-loss policies:
    \[\mathcal{I}_{3}:=\{I(x)=\lambda(x-d)_+; d\in\overline{\mathbb{R}}_+,\lambda\in[0,1]\}.\]
\end{itemize}
It is remarked that any $I(\cdot)$ in $\mathcal{I}_{1}\cup \mathcal{I}_{2}\cup \mathcal{I}_{3}$ is increasing, convex, and 1-Lipschitz.
For $i=1,2,3$, we denote by $\mathcal{M}_i$ the collection of all feasible menus $M=\{(I_{(a,k)},P_{(a,k)})\mid (a,k)\in \mathbb{R}_+\times K\}$ with $I_{(a,k)}(\cdot)\in \mathcal{I}_i$ for all $(a,k)$. We also write
$\mathcal{M}:=\mathcal{M}_1\cup\mathcal{M}_2\cup\mathcal{M}_3$.

To sum up, the following three problems will be considered in this paper:
\begin{equation}\label{M123}
\max_{M\in\mathcal{M}_i}\left\{J[M]=\int_{\mathbb{R}_+\times K}(P_{(a,k)} - H[I_{(a,k)}(X_k)])\,{\mathbb{Q}}(da\times dk)\right\},\quad i=1,2,3.
\end{equation}
Implicit in this formulation is that we only consider those feasible menus $M$ that make $J[M]$ is well-defined, that is, the integrand is both measurable and integrable. We further impose the following integrability condition:
\begin{equation}\label{Integrability}
\int_{\mathcal{T}}\bigl(a+H[X_k]\bigr)\,\mathbb{Q}(da\times dk)<\infty,
\end{equation}
which is a mild technical condition to ensure the well-posedness of the optimization problems below and the existence of the relevant optimal solutions.

\section{The failure of first-best solutions}\label{sec:failure}
This short section is devoted to explain why first-best solutions do not work by juxtaposing the situations with and without complete information about the true types of agents.

Under complete information, the type $(\alpha,k)$ (or $(a,k)$) of each agent (insurance companies) is known to the principal (the reinsurer) and is not hidden. In this case, the reinsurer can design individual tailor-made policy for each agent so as to maximize her expected profit without referencing the distribution of types $\bQ$. The policy $(I_{(a,k)},P_{(a,k)})$ to be offered to agent of type $(a,k)$ must satisfy the  IR constraint (\ref{IR-1}) in order for it to be accepted by the agent, that is,
simplifying the condition  $P_{(a,k)}\leq I_{(a,k)}(a)$.
Therefore, regardless of the functional form of the indemnity $I_{(a,k)}$, the premium $P_{(a,k)}$ must equal  $I_{(a,k)}(a)$, which is a function of $I_{(a,k)}$. The principal chooses the optimal indemnity $I^{FB}_{(a,k)}$ to maximize her profit:
\[
P_{(a,k)} - H[ I_{(a,k)}(X_k)]=I_{(a,k)}(a)- H[ I_{(a,k)}(X_k)],
\]
and the policy $(I^{FB}_{(a,k)},P^{FB}_{(a,k)})$ with $P^{FB}_{(a,k)} = I^{FB}_{(a,k)}(a)$ is called the {\em first-best} policy.

Let us consider what would happen when there is
information asymmetry so that the reinsurer only knows the distribution of types but not the exact type of the agent (say $(a_1,k_1)$). If the reinsurer continues to offer  first-best policies $\{(I^{FB}_{(a,k)},P^{FB}_{(a,k)})\mid (a,k)\in \cl{T}\}$ to the pool of agents, will agents truthfully reveal their hidden types by self-selecting the policies tailor-made for them, so that the reinsurer can earn the same profit as in the case of complete information?

If agent  $(a_1,k_1)$ (or equivalently $(\alpha_1,k_1)$) picks $(I^{FB}_{(a_1,k_1)},P^{FB}_{(a_1,k_1)})$, his risk reduction is
\[
\text{V$@$R}_{\alpha_1} (X_{k_1})-  \text{V$@$R}_{\alpha_1} (X_{k_1} - I^{FB}_{(a_1,k_1)}(X_{k_1}) +P^{FB}_{(a_1,k_1)}) = I^{FB}_{(a_1,k_1)}(a_1)-P^{FB}_{(a_1,k_1)}=0
\]
However, if this agent  picks the policy $(I^{FB}_{(a_2,k_2)},P^{FB}_{(a_2,k_2)})$ with $a_2<a_1$, his risk reduction becomes
\begin{align*}
\text{V$@$R}_{\alpha_1} (X_{k_1})-  \text{V$@$R}_{\alpha_1} (X_{k_1} - {I^{FB}_{(a_2,k_2)}}(X_{k_1}) +{P^{FB}_{(a_2,k_2)}})
&=I^{FB}_{(a_2,k_2)}(a_1)-P^{FB}_{(a_2,k_2)}\\
&=I^{FB}_{(a_2,k_2)}(a_1)-I^{FB}_{(a_2,k_2)}(a_2)\\
&\geq 0.
\end{align*}
Therefore, self-selection will not happen: the agent of type $(a_1,k_1)$ is better off by {\em mimicking} to be type $(a_2,k_2)$ with $a_2<a_1$. This is feasible as his true type is hidden and is allowed to choose any policy available in the menu. When this happens,  the profit of the reinsurer is no longer the first-best profit
$P^{FB}_{(a_1,k_1)} - H[I^{FB}_{(a_1,k_1)}(X_{k_1})]$,
it becomes
\begin{align*}
 P^{FB}_{(a_2,k_2)} - H[I^{FB}_{(a_2,k_2)}(X_{k_1})] & = I^{FB}_{(a_2,k_2)}(a_2) - H[I^{FB}_{(a_2,k_2)}(X_{k_1})]\\
  &\leq I^{FB}_{(a_2,k_2)}(a_1) - H[I^{FB}_{(a_2,k_2)}(X_{k_1})]\\
  &\leq I^{FB}_{(a_1,k_1)}(a_1) - H[I^{FB}_{(a_1,k_1)}(X_{k_1})]
\end{align*}
because $I^{FB}_{(a_1,k_1)}$ maximizes
$I(a_1) - H[I(X_{k_1})]$
over $I$  by definition. We conclude that first-best profit is never achievable when the agent has hidden information of his own characteristic.

Motivated by the failure of first-best policies, we adopt the  PA model to look for {\em second-best} policies. It is a framework to elicit agent's hidden information by optimally giving up some (information) rent to the agents. This is achieved by imposing IC constraint to induce agents to reveal their hidden characteristics by self-selecting the tailor-made polices, with a tradeoff of giving up part of the first-best profit so that agents can enjoy a higher level of risk reduction.

\section{Indirect utility}\label{sec:indirect}
The problems in (\ref{M123}) are infinite-dimensional and generally non-convex, involving infinitely many premium variables $P_{(a,k)}$ with no a priori functional form, together with globally coupled incentive-compatibility constraints. Our first key step is therefore to introduce the notion of \textit{indirect utility}, following the established literature on adverse-selection problems; see for example  \cite{lester2019screening}  and \cite{bajari2014moral}. We exploit the specific structure of V$@$R, together with the change of variables introduced above, to represent the relevant indirect utility by the univariate function $v_M(a)$ defined below. This reduction replaces the infinitely many premium variables with a single convex function, captures the relationship between the optimal premiums $P_{(a,k)}^*$ and indemnities $I_{(a,k)}^*$, and, as shown below, substantially simplifies the incentive-compatibility constraints.\footnote{A related indirect-utility formulation appears in \cite{sung2012optimal}. The present analysis develops this idea for continuum-type reinsurance menus and uses the convexity and subgradient structure of $v_M$ to characterize the optimal contracts.}

Let $M=\{(I_{(a,k)},P_{(a,k)})\mid (a,k)\in \mathbb{R}_+\times K\}\in\cl{M}$ be a  feasible menu. We associate to $M$ an indirect utility function $v_M:\mathbb{R}_+\to\mathbb{R}$ which  is defined as
\begin{equation}\label{IndirectUtility}
v_{M}(a) := \sup_{\left(I_{(a',k')},P_{(a',k')}\right)\in M}\{I_{(a',k')}(a)-P_{(a',k')}\},\quad a\in\mathbb{R}_+.
\end{equation}
Note that
\begin{align*}
    \text{V$@$R}_\alpha (X_k)-\left(\text{V$@$R}_\alpha (X_k - I_{(\alpha',k')}(X_k) )+P_{(\alpha',k')}\right) & = I_{(\alpha',k')}(\text{V$@$R}_\alpha(X_k)) -P_{(\alpha',k')}\\
    & = I_{(\alpha',k')}(a) -P_{(\alpha',k')},
\end{align*}
where $I_{(a',k')}(a)-P_{(a',k')}$ is the amount of risk reduction if an agent of type $(a,k)$ chooses contract $(I_{(a',k')},P_{(a',k')})$. As a result, $v_M(a)$ is simply the largest possible risk reduction an agent of type $(a,k)$ can benefit when the menu $M$ is offered.

The next lemma summarises the essential properties of $v_M$.

\begin{lemma}\label{L:IndirectUtility}
For any feasible menu $M=\{(I_{(a,k)},P_{(a,k)})\mid (a,k)\in \mathbb{R}_+\times K\}\in\mathcal{M}$, the indirect utility function $v_M$ is increasing, convex, and 1-Lipschitz on $\mathbb{R}_+$, and is given by
\begin{equation}\label{v}
v_M(a)=I_{(a,k)}(a)-P_{(a,k)},
\end{equation}
in which the right hand side is independent of $k\in K$. In particular,
\[(v_M)_+'(+\infty)\leq 1.\]
\end{lemma}
{\em Proof.} Fix a feasible menu $M=\{(I_{(a,k)},P_{(a,k)})\mid (a,k)\in \mathbb{R}_+\times K\}\in\mathcal{M}$. For any $(a',k')\in \mathbb{R}_+\times K$, the function $a\mapsto I_{(a',k')}(a)-P_{(a',k')}$ is increasing and convex, so the supremum $a\mapsto v_M(a)$ is also increasing and convex.
For any $(a,k)\in \mathbb{R}_+\times K$, IC constraint says that
\[I_{(a,k)}(a)-P_{(a,k)}\geq I_{(a',k')}(a)-P_{(a',k')}\quad\forall (a',k')\in\mathbb{R}_+\times K,\]
it is clear that the supremum in (\ref{IndirectUtility}) is achieved at $(a,k)$ for any $k\in K$, and hence
\[v_M(a)=I_{(a,k)}(a)-P_{(a,k)}.\]
To prove that $v_M$ is 1-Lipschitz, we assume that for some $0\leq a_1<a_2$, $v_M(a_2)-v_M(a_1)>a_2-a_1$. By definition,
\[v_M(a_1)= \sup_{(I_{(a',k')},P_{(a',k')})\in M}\{I_{(a',k')}(a_1)-P_{(a',k')}\}\geq I_{(a_2,k)}(a_1)-P_{(a_2,k)},\]
for any $k\in K$, hence
\begin{align*}
    I_{(a_2,k)}(a_2)-I_{(a_2,k)}(a_1)&=(I_{(a_2,k)}(a_2)-P_{(a_2,k)})-(I_{(a_2,k)}(a_1)-P_{(a_2,k)})\\
    &=v_M(a_2)-(I_{(a_2,k)}(a_1)-P_{(a_2,k)})\\
    &\geq v_M(a_2)-v_M(a_1)\\
    &>a_2-a_1,
\end{align*}
which is impossible as $I_{(a,k)}$ is 1-Lipschitz for any $(a,k)$. This proves that $v_M$ is 1-Lipschitz, which finishes the proof.\hfill $\square$

Lemma \ref{L:IndirectUtility} states that for any feasible menu $M=\{(I_{(a,k)},P_{(a,k)})\mid (a,k)\in \mathbb{R}_+\times K\}\in\mathcal{M}$, the map
\begin{equation}\label{Prem-v}
a\mapsto v_M(a)=I_{(a,k)}(a) - P_{(a,k)}
\end{equation}
is independent of the choice of $k$ on the right hand side, and it is necessary that the map $v_M$ is increasing, convex, and 1-Lipschitz.
While $v_M$ is only a function of $a$ and is not depending on $k$ directly,  one should bear in mind that $a=\text{V$@$R}_\alpha (X_k)$ is indeed depending on $k$. Lemma \ref{L:IndirectUtility} suggests an important phenomenon that regardless of the menu to be offered, agents will a larger risk exposure, as measure by $a=\text{V$@$R}_\alpha (X_k)$, can always benefit more than those with a smaller risk exposure, measured in terms of the amount of risk reduction upon purchasing the tailor-made policies in the menu.

Another implication of Lemma \ref{L:IndirectUtility}, and relationship (\ref{v}) in particular, is that the menu, which is originally represented by a collection of policies  $(I_{(a,k)},P_{(a,k)})$ indexed by $(a,k)\in \mathbb{R}_+\times K$, can  be equivalently represented by just a collection of indemnity functions $I_{(a,k)}$ indexed by $(a,k)\in \mathbb{R}_+\times K$ together with the indirect utility function $v_M$. We schematically represent this important observation by
\[M=\{(I_{(a,k)},P_{(a,k)})\mid (a,k)\in \mathbb{R}_+\times K\}\longleftrightarrow\{v_M,I_{(a,k)}\mid (a,k)\in \mathbb{R}_+\times K\}.\]
Because of (\ref{v}), the objective function in (\ref{M123}) becomes
\begin{equation}\label{J0}
J[M]:=\int\{I_{(a,k)}(a) - H[I_{(a,k)}(X_k)]-v_M(a)\}\mathbb{Q}(da\times dk),
\end{equation}
 while the IC constraints become
\begin{equation}\label{IC0}
v_M(a')-v_M(a)\geq I_{(a,k)}(a')-I_{(a,k)}(a),\quad \text{for any }(a,k)\in\bR_+\times K\text{ and }a'\in\bR_+,
\end{equation}
and the IR constraints become
\begin{equation}\label{IR0}
v_M(a)\geq 0,\quad a\in\mathbb{R}_+.
\end{equation}
Note that under this new formulation, the decision variables $P_{(a,k)}$ are replaced by the function $v_M$ which is analytically much tractable.

It is important to observe that the IC constraints can be expressed as a collection of constraints indexed by $(a,k)\in\bR_+\times K$:
\[\text{IC~(\ref{IC-1})} \Longleftrightarrow \{\text{IC}(a,k):(a,k)\in\bR_+\times K\},\]
where $\text{IC}(a,k)$ is the constraint
\[\text{IC}(a,k):\qquad v_M(a')-v_M(a)\geq I_{(a,k)}(a')-I_{(a,k)}(a),\quad \text{for any }a'\in\mathbb{R}_+.\]
In the IC constraint (\ref{IC-1}) under the original formulation (without the use of indirect utility), all indemnities $I_{(a,k)}$ intervene and couple with each other, making analysis and verification extremely difficult. The merit of introducing the notion of indirect utility is that it greatly simplifies the situation by ``decoupling'' the IC constraint  into a collection of $\text{IC}(a,k)$ constraints in which each such $\text{IC}(a,k)$ constraint only involves one indemnity, namely $I_{(a,k)}$, but not others. Another merit is that via (\ref{Prem-v}), we can encode all $P_{(a,k)}$ whose structure is yet not known by a single indirect utility function  which is shown to have tractable mathematical properties (such as increasing and convex).
In subsequent analysis, at times we consider the case when $v$ is fixed and treat
the $\text{IC}(a,k)$ as a constraint on $I_{(a,k)}$. In this case, we denote such constraint as $\text{IC}_v(a,k)$.

From (\ref{IC0}) and (\ref{IR0}), it is evident that translating the indirect utility function $v_M$ by an additive constant does not affect the IR and IC constraints as long as $v_M$ remains positive.  As $J[M]$ in (\ref{J0}) is decreasing in $v_M$, we conclude that one must have $v_M(0)=0$ at optimal.
From now on, we denote by $\cl{V}$ the collection of all increasing, convex, and 1-Lipschitz functions on $\mathbb{R}_+$ that vanish at 0. All $v\in\mathcal{V}$ automatically satisfy the IR constraints.

With the introduction of indirect utility functions and simplifications described above, the three problems in (\ref{M123}) are now reduced to
\begin{equation}\label{J1}
\max_{M\in\mathcal{M}_i}\left\{J[M]=\int_{\mathbb{R}_+\times K}\{I_{(a,k)}(a) - H[I_{(a,k)}(X_k)]-v(a)\}\,{\mathbb{Q}}(da\times dk)\right\},\quad i=1,2,3,
\end{equation}
where we simply write $v_M(\cdot)$ as $v(\cdot)$ with a slight abuse of notation whenever the feasible menu set $M$ is fixed, and $\mathcal{M}_i$ is the collection of all menus
\[M=\{v, I_{(a,k)}\mid v\in\mathcal{V}, I_{(a,k)}\in\mathcal{I}_i\text{ for all }(a,k)\in\bR_+\times K \}\] satisfying the IC constraints: for any $(a,k)\in\bR_+\times K$:
\[v(a')-v(a)\geq I_{(a,k)}(a')-I_{(a,k)}(a),\quad \text{for any }a'\in\mathbb{R}_+.\]
These three problems will be analyzed and solved completely one by one in the next three sections.

\section{Class of stop-loss policies}\label{sec:stoploss}

In this section, we consider the class of feasible menus from
$\mathcal{M}_1$.
For any such menu
\[M=\{v, I_{(a,k)}\mid v\in\mathcal{V}, I_{(a,k)}\in\mathcal{I}_1\text{ for all }(a,k)\in\bR_+\times K \}\in\mathcal{M}_1,\]
we have
$I_{(a,k)}(x)=(x-d_{(a,k)})_+$ for all $(a,k)$. For notational simplicity, we write such menu as
$M=\{v, d_{(a,k)}\}$.
For any feasible menu $M=\{v, d_{(a,k)}\}\in\mathcal{M}_1$,
the function $v\in\mathcal{V}$ and the numbers $d_{(a,k)}\in\overline{\mathbb{R}}_{+}$ must satisfy the IC$(a,k)$ constraint
\[
v(a')-v(a)\geq (a'-d_{(a,k)})_+-(a-d_{(a,k)})_+,\quad \text{for any }a'\in\mathbb{R}_+,
\]
for any $(a,k)\in\mathbb{R}_+\times K$. Substituting the specific form of the indemnity functions in (\ref{J1}), we can formally state the problem we aim to solve in this section as follows:
\begin{equation}\label{Problem1}
\max_{M\in\mathcal{M}_1}\left\{J[M]=\int_{\mathbb{R}_+\times K}\{(a-d_{(a,k)})_+ - H[(X_k-d_{(a,k)})_+]-v(a)\}\,{\mathbb{Q}}(da\times dk)\right\}.
\end{equation}

The next lemma is crucial in that it gives us necessary conditions that optimal $d_{(a,k)}$ must satisfy, and it also identifies the optimal shape of $v$.

\begin{lemma}\label{SL-v}
Let  $M=\{v, d_{(a,k)}\}\in\mathcal{M}_1$ be a feasible menu. Then

\noindent (i)
for any $k\in K$,
\begin{equation}\label{v'}
v_-'(a)\leq \mathbf{1}_{\{a>d_{(a,k)}\}}\leq \mathbf{1}_{\{a\geq d_{(a,k)}\}}\leq v_+'(a)\quad\text{for all $a> 0$},
\end{equation}
and
\begin{equation}\label{v'2}
 \mathbf{1}_{\{0\geq d_{(0,k)}\}}\leq v_+'(0);
\end{equation}
(ii) the function $v$ must take the form
$$v(a)=(a-\tau)_+,\quad a\in\bR_+,$$
for some $\tau\in\overline{\mathbb{R}}_{+}$.
\end{lemma}
{\em Proof.} (i) Fix any $(a,k)\in\mathbb{R}_+\times K$. Let $g(x):=(x-d_{(a,k)})_+$. The right-hand derivative is $g_+'(x)= \mathbf{1}_{\{x\geq d_{(a,k)}\}}$. Now consider $a'\downarrow a$, we have
\[v_+'(a)=\lim_{a'\downarrow a}\frac{v(a')-v(a)}{a'-a}\geq \lim_{a'\downarrow a}\frac{g(a')-g(a)}{a'-a}=g_+'(a)=\mathbf{1}_{\{a\geq d_{(a,k)}\}}.\]
Similarly, for $a>0$, we can obtain
\[
v_-'(a)=\lim_{a'\uparrow a}\frac{v(a)-v(a')}{a-a'}\leq \lim_{a'\uparrow a}\frac{g(a)-g(a')}{a-a'}=g_-'(a)= \mathbf{1}_{\{a>d_{(a,k)}\}}.
\]
Combining these inequalities gives (\ref{v'}) and (\ref{v'2}).

(ii) Being a convex function, $v$ is differentiable everywhere on $\mathbb{R}_+$ except at countably many points. Thus for all $a\in(0,\infty)\setminus\mathcal{N}$ for some countable set $\mathcal{N}$, $v'(a)=v_-'(a)=v_+'(a)$, and so from part (i) we obtain \[v'(a)=\mathbf{1}_{\{a>d_{(a,k)}\}}=\mathbf{1}_{\{a\geq d_{(a,k)}\}},\quad a\in(0,\infty)\setminus\mathcal{N}.\]
Let $\tau:=\sup\{a: v(a)=0\}$. Since $v'$ is either 1 or 0 except at countably many points, $v$ must take the form
\[
v(a)=(a-\tau)_+,\quad a\in\bR_+,
\]
for some $\tau\in\overline{\mathbb{R}}_{+}$. \hfill $\square$

From Lemma \ref{SL-v}, if $M=\{v, d_{(a,k)}\}\in\mathcal{M}_1$, then $v$ must take the form
$v(a)=(a-\tau)_+$,
for some $\tau\in\overline{\mathbb{R}}_{+}$. Thus we may express $v\in\cl{V}$ be the single number $\tau\in\overline{\mathbb{R}}_{+}$, and hence the menu $M$ can be equivalently represented as $M=\{\tau, d_{(a,k)}\}$.
This way of representing a feasible menu will be adopted in the reminder of this section. It also follows from Lemma \ref{SL-v}
that
\begin{equation}\label{a1}
 a\in[0,\tau) \Longrightarrow v'(a)=0,  \Longrightarrow a<d_{(a,k)}
\end{equation}
and
\begin{equation}\label{a2}
a\in(\tau,\infty) \Longrightarrow  v'(a)=1 \Longrightarrow a> d_{(a,k)}.
\end{equation}
When $a=\tau$, $v'_-(\tau)=0$  and $v'_+(\tau)=1$, so $d_{(\tau,k)}$ is not constrained by (\ref{v'}).

By making use of Lemma \ref{SL-v}, we can simplify the IC constraints as follows:

\begin{lemma}\label{SL-IC} Consider the menu
$M=\{v, d_{(a,k)}\}$
with $v(\cdot)=(\cdot-\tau)_+\in\mathcal{V}$ for some fixed $\tau\in\overline{\mathbb{R}}_{+}$. For any $(a,k)\in \cl{T}$,
\begin{equation}\label{SL-IC2}
\mathrm{IC}_v(a,k)\Longleftrightarrow
\begin{cases}
d_{(a,k)}\geq\tau, & \text{if } a<\tau,\\
d_{(a,k)} \in\overline{\mathbb{R}}_{+}, & \text{if } a=\tau,\\
d_{(a,k)}\leq\tau, & \text{if } a>\tau.
\end{cases}
\end{equation}
\end{lemma}
{\em Proof.} First consider $a<\tau$. Assume that $\text{IC}_v(a,k)$ holds. From (\ref{a1}), $a<d_{(a,k)}$. Hence
\begin{align*}
\text{IC}_v(a,k)& \Longleftrightarrow   v(a')-v(a)\geq (a'-d_{(a,k)})_+-(a-d_{(a,k)})_+,\quad \text{for any }a'\in\mathbb{R}_+\\
&\Longrightarrow   (a'-\tau)_+\geq (a'-d_{(a,k)})_+,\quad \text{for any }a'\in\mathbb{R}_+\\
&\Longleftrightarrow d_{(a,k)}\geq\tau.
\end{align*}
Conversely,
\begin{align*}
 d_{(a,k)}\geq\tau &\Longleftrightarrow  (a'-\tau)_+\geq (a'-d_{(a,k)})_+,\quad \text{for any }a'\in\mathbb{R}_+\\
&\Longrightarrow  (a'-\tau)_+-0\geq (a'-d_{(a,k)})_+-(a-d_{(a,k)})_+,\quad \text{for any }a'\in\mathbb{R}_+\\
&\Longleftrightarrow  (a'-\tau)_+-v(a)\geq (a'-d_{(a,k)})_+-(a-d_{(a,k)})_+,\quad \text{for any }a'\in\mathbb{R}_+\\
&\Longleftrightarrow \text{IC}_v(a,k).
\end{align*}
This proves that IC${}_v(a,k)\Longleftrightarrow d_{(a,k)}\geq\tau$ when $a<\tau$.

Next, we consider the case when $a=\tau$. In this case,
\begin{align*}
\text{IC}_v(\tau,k)& \Longleftrightarrow   v(a')-v(\tau)\geq (a'-d_{(\tau,k)})_+-(\tau-d_{(\tau,k)})_+,\quad \text{for any }a'\in\mathbb{R}_+\\
&\Longleftrightarrow   (a'-\tau)_+\geq (a'-d_{(\tau,k)})_+-(\tau-d_{(\tau,k)})_+,\quad \text{for any }a'\in\mathbb{R}_+,
\end{align*}
which always holds true for any $d_{(\tau,k)}\in\overline{\mathbb{R}}_{+}$.

Finally, we consider the case when $a>\tau$. Assume that $\text{IC}_v(a,k)$ holds. From (\ref{a2}), $d_{(a,k)}<a$. Hence
\begin{align*}
\text{IC}_v(a,k)& \Longleftrightarrow   v(a')-v(a)\geq (a'-d_{(a,k)})_+-(a-d_{(a,k)})_+,\quad \text{for any }a'\in\mathbb{R}_+\\
&\Longrightarrow   (a'-\tau)_+-(a-\tau)\geq (a'-d_{(a,k)})_+-(a-d_{(a,k)}),\quad \text{for any }a'\in\mathbb{R}_+\\
&\Longleftrightarrow   (a'-\tau)_++\tau\geq (a'-d_{(a,k)})_++d_{(a,k)},\quad \text{for any }a'\in\mathbb{R}_+\\
&\Longleftrightarrow \tau\geq d_{(a,k)}.
\end{align*}
Conversely,
\begin{align*}
\tau \geq d_{(a,k)}&\Longleftrightarrow  (a'-\tau)_++\tau\geq (a'-d_{(a,k)})_++d_{(a,k)},\quad \text{for any }a'\in\mathbb{R}_+\\
&\Longleftrightarrow  (a'-\tau)_+-v(a)\geq (a'-d_{(a,k)})_+-(a-d_{(a,k)}),\quad \text{for any }a'\in\mathbb{R}_+\\
&\Longrightarrow  (a'-\tau)_+-v(a)\geq (a'-d_{(a,k)})_+-(a-d_{(a,k)})_+,\quad \text{for any }a'\in\mathbb{R}_+\\
&\Longleftrightarrow \text{IC}_v(a,k).
\end{align*}
This proves that IC${}_v(a,k)\Longleftrightarrow d_{(a,k)}\leq\tau$ when $a>\tau$.\hfill $\square$

From Lemma \ref{SL-v} and Lemma \ref{SL-IC}, the problem we want to solve is to maximize $J[M]$ in (\ref{Problem1}) over all  menus $M=\{\tau,d_{(a,k)}\}$ that satisfy the IC$(a,k)$ constraint (\ref{SL-IC2}) for all $(a,k)$. The next result identifies the optimal values of $d_{(a,k)}$ for all $(a,k)\in\cl{T}$ when $\tau$ is fixed.

\begin{theorem}\label{T-SL}
For any $\tau\in\overline{\mathbb{R}}_{+}$ and $(a,k)\in\cl{T}$, define
\begin{equation}\label{OptimalD}
d^*_{(a,k)}(\tau):=
\begin{cases}
\theta_k^*\wedge \tau, & a>\tau, \\
\theta_k^*, & a=\tau\geq \xi_k,\\
+\infty, & a=\tau<\xi_k,\\
+\infty, & a<\tau.
\end{cases}
\end{equation}
and
\begin{equation}\label{Phi}
    \Phi_{(a,k)}(\tau) := (a-d_{(a,k)}^*(\tau))_+ - H[(X_k-d_{(a,k)}^*(\tau))_+]-(a-\tau)_+.
\end{equation}
There is an optimal solution $\tau^*\in\overline{\mathbb{R}}_{+}$ maximizing the reinsurer's profit
\begin{equation}\label{D-tau}
\int_{\cl{T}}\Phi_{(a,k)}(\tau)\mathbb{Q}(da\times dk),
\end{equation}
and the optimal deductibles and optimal premiums for Problem (\ref{Problem1}) are given by $d^*_{(a,k)}(\tau^*)$ and
\begin{equation}\label{OptimalD-P}
P^*_{(a,k)}=
\begin{cases}
\tau^*-\theta_k^*\wedge \tau^*, & a>\tau^*,\\
\tau^*-\theta_k^*, & a=\tau^*\geq \xi_k,\\
0, & a=\tau^*<\xi_k,\\
0, & a<\tau^*,\\
\end{cases}
\end{equation}
respectively.
\end{theorem}
{\em Proof.} The first step of the proof is to demonstrate that for any fixed $\tau\in\overline{\mathbb{R}}_+$ and  $(a,k)\in\cl{T}$, the integrand of the objective function in (\ref{Problem1}) is maximized pointwise when $d_{(a,k)}$ equals $d_{(a,k)}^*(\tau)$ given by (\ref{OptimalD}).To this end , we first consider the case when $(a,k)\in\cl{T}$ with $a<\tau$. From (\ref{SL-IC2}), $d_{(a,k)}$ has to satisfy the IC${}_v(a,k)$ constraint $d_{(a,k)}\geq\tau$, and hence the integrand in (\ref{Problem1}) becomes
\[
(a-d_{(a,k)})_+ - H[(X_k-d_{(a,k)})_+]-v(a)\\
=- H[(X_k-d_{(a,k)})_+].
\]
We obtain $d_{(a,k)}^*(\tau)$ by maximizing the expression above, subject to $d_{(a,k)}\geq\tau$. The monotonicity of $H$ leads to $d^*_{(a,k)}(\tau)=+\infty$, and the maximum value of the integrand is zero.

Next, we consider the case when $(a,k)\in\cl{T}$ with $a=\tau$. This time, $d_{(a,k)}$ is unconstrained by (\ref{SL-IC2}), and the integrand in (\ref{Problem1}) equals
\begin{align*}
(\tau-d_{(\tau,k)})_+ - H[I_{(\tau,k)}(X_k)]-v(\tau)
= (\tau-d_{(\tau,k)})_+ - H[(X_k-d_{(\tau,k)})_+].
\end{align*}
We are led to the following maximization problem:
\[
\max_{d_{(\tau,k)}\geq 0}\left[(\tau-d_{(\tau,k)})_+ -(1+\theta)\int_{d_{(\tau,k)}}^\infty (h\circ \overline{F}_k)(y)\,dy\right].
\]
It is straightforward to solve this problem  and obtain:
\[
d^*_{(\tau,k)}(\tau)=
\begin{cases}
+\infty, & \text{if $\tau<\xi_k $,}\\
\theta^*_k,  & \text{if $\tau\geq\xi_k$}.
\end{cases}
\]

The final case to consider is $(a,k)\in\cl{T}$ with $a>\tau$, From (\ref{SL-IC2}), $d_{(a,k)}$ should satisfy the IC${}_v(a,k)$ constraint $d_{(a,k)}\leq\tau$, and hence the integrand in (\ref{Problem1}) equals
\[
(a-d_{(a,k)})_+ - H[I_{(a,k)}(X_k)]-v(a)=a-d_{(a,k)} - H[(X_k-d_{(a,k)})_+]-v(a)
\]
Maximizing the expression on the right subject to $0\leq d_{(a,k)}\leq\tau$ yields
$d^*_{(a,k)}(\tau)=\theta^*_k\wedge \tau$.

Combining all cases above gives (\ref{OptimalD}), the  optimal value of the deductible $d_{(a,k)}^*(\tau)$ for all $(a,k)\in\cl{T}$ as a function of $\tau\in\overline{\mathbb{R}}_+$. Upon substituting $d_{(a,k)}^*(\tau)$ in the integrand in (\ref{Problem1}), the expected profit of the reinsurer is then given by the expression in (\ref{D-tau}), which is a function of $\tau\in[0,\infty]$. To obtain the optimal menu, it remains to find $\tau^*$ that maximizes (\ref{D-tau}); the optimal deductibles are then given by $d_{(a,k)}^*(\tau^*)$, while the optimal premiums can then be obtained from equation (\ref{v}) in Lemma \ref{L:IndirectUtility}:
\[
P_{(a,k)}^*=I_{(a,k)}^*(a)-v_{M^*}(a)=(a-d_{(a,k)}^*(\tau^*))_+-(a-\tau^*)_+,
\]
which is just the expression  in (\ref{OptimalD-P}) after simplification.

To complete the proof, it remains to show the existence of an optimizer of (\ref{D-tau}) on $\overline{\mathbb{R}}_+$. This  can be proved by first observing the upper semicontinuity of the map $\tau\mapsto \Phi_{(a,k)}(\tau)$ for any $(a,k)$, and then using reverse Fatou's lemma.
\hfill $\square$

In Theorem \ref{T-SL}, $\tau^*$ is defined as an maximizer of
(\ref{D-tau}). While closed form expression of $\tau^*$ is not available in general, we nevertheless have the following result.

\begin{proposition}\label{L:tau>mina}
$\tau^*$ defined as an maximizer of  (\ref{D-tau}) satisfies
\[
\tau^*\geq L:= \inf_{(\alpha,k)\in\tilde{\cl{T}}}\text{V$@$R}_\alpha (X_k) =\inf \{a\in\bR_+: \exists k\in K \text{ such that }(a,k)\in\cl{T}\}.
\]
\end{proposition}
{\em Proof.} Suppose that $\tau<L$. According to Theorem \ref{T-SL}, $a>\tau$ for any $(a,k)\in \cl{T}$, $d^*_{(a,k)}(\tau)=\theta_k^*\wedge\tau$ and $P^*_{(a,k)}(\tau):=(a-d^*_{(a,k)}(\tau))_+-(a-\tau)_+=\tau-\theta_k^*\wedge \tau$, and hence $\Phi_{(a,k)}(\tau) $ in (\ref{Phi}) simplifies to
\[
    \Phi_{(a,k)}(\tau) = \tau-\theta_k^*\wedge\tau - H[(X_k-\theta_k^*\wedge\tau)_+].
\]
This formula for $\Phi_{(a,k)}(\tau) $ holds true whenever $\tau<L$; moreover, it is increasing in $\tau$. Thus the objective function
in problem (\ref{D-tau}) is increasing in $\tau$ when $\tau<L$, and hence any $\tau<L$ cannot be optimal.\hfill $\square$

The following numerical example illustrates the theoretical findings in Theorem \ref{T-SL}.
\begin{example}\label{EG:ST}
	Suppose that an agent of type $(\alpha,k)$ faces an exponentially distributed loss $X_k$ with mean $k$, where $k\sim\mathrm{U}(5000,25000)$ and $\alpha\sim\mathrm{U}(0.01,0.025)$ are independent. Here $\alpha$ is a tail probability, so this interval corresponds to conventional confidence levels between $97.5\%$ and $99\%$. Let $H$ be given by (\ref{H:Example}) with $\theta=0.1$ and the linear--quadratic distortion $h_\gamma(x)=(1+\gamma)x-\gamma x^2$, $\gamma\in[0,1]$. Since $\overline F_k(x)=\exp(-x/k)$,
\[
H[(X_k-d)_+]
=1.1k\left[(1+\gamma)\exp\left(-\frac{d}{k}\right)
-\frac{\gamma}{2}\exp\left(-\frac{2d}{k}\right)\right].
\]
The transformed type is $a=\text{V$@$R}_\alpha(X_k)=-k\log\alpha$. Writing
$\underline c=-\log(0.025)\approx3.688879$ and
$\overline c=-\log(0.01)\approx4.605170$, the conditional distribution of $a$ has support $[\underline c k,\overline c k]$ and density
\[
f(a;k)=\frac{1}{0.015k}\exp\left(-\frac{a}{k}\right)
\mathbf{1}_{\{\underline c k\leq a\leq\overline c k\}}.
\]

For the baseline $\gamma=0.5$, direct calculation gives
$\theta_k^*\approx0.170941k$ and $\xi_k\approx1.366310k$.
By Lemma \ref{SL-v}, $v(a)=(a-\tau)_+$. Moreover,
$L=5000\underline c\approx18444.40$, while
$\sup_k\theta_k^*\approx4273.53$. Since $L>\sup_k\theta_k^*$, Proposition  \ref{L:tau>mina} reduces (\ref{D-tau}) to
\[
\max_{\tau\geq L}\frac{1}{20000}\int_{5000}^{25000}
(\tau-1.366310k)\Pr(a>\tau\mid k)\,dk,
\]
with numerical solution $\tau^*\approx61087.34$ and $J[M^*]\approx16831.00$. Equation (\ref{OptimalD-P}) then yields
\[
d^*_{(a,k)}=
\begin{cases}
0.170941k, & a\geq61087.34,\\
+\infty, & a<61087.34,
\end{cases}
\]
and
\[
P^*_{(a,k)}=
\begin{cases}
61087.34-0.170941k, & a\geq61087.34,\\
0, & a<61087.34.
\end{cases}
\]

Under full information, every type is served because $a\geq\underline c k>\xi_k$. Under asymmetric information, by contrast, only $49.80\%$ of agents participate: average profit falls from its first-best level of $40675.64$ to $16831.00$, or $41.38\%$ of first-best profit, while the population-average information rent is $10327.74$. Thus the example exhibits directly the information-rent--market-exclusion tradeoff underlying Theorem \ref{T-SL}.

To isolate the effect of heterogeneity in $\alpha$, set
$\alpha_0=\exp(\mathbb{E}[\log\alpha])\approx0.016941$. This choice preserves
$\mathbb{E}[-\log\alpha]$, and hence the average first-best surplus, while making
$a=c_0k$, where $c_0=-\log\alpha_0\approx4.078019$. The optimization becomes
\[
\max_{5000c_0\leq\tau\leq25000c_0}\frac{1}{20000}\int_{\tau/c_0}^{25000}
(\tau-1.366310k)\,dk,
\]
and therefore
\[
\tau_0^*=\frac{25000c_0^2}{2c_0-1.366310}\approx61233.08,
\qquad
k_0^*=\frac{\tau_0^*}{c_0}\approx15015.40.
\]
The deductible remains $0.170941k$, but the contract is now offered precisely when
$k\geq k_0^*$, at premium $P_k^*=\tau_0^*-0.170941k$.

Table \ref{tab:SL-alpha-comparison} compares the heterogeneous and degenerate tail probabilities. Because the mean transformed tail index $\mathbb{E}[-\log\alpha]$, and hence mean willingness to pay, is matched, the two cases have the same first-best profit. Heterogeneity in $\alpha$ raises information rent by about $1.6\%$ and lowers optimal profit by about $0.5\%$, while barely changing aggregate participation. Its main qualitative effect is compositional: with fixed $\alpha$, participation is determined by a single cutoff in $k$; with heterogeneous $\alpha$, agents sharing the same loss distribution may differ in participation because of their tail probabilities, although the contract offered conditional on participation remains $k$-dependent.
\begin{table}[H]
\centering
\caption{Heterogeneous versus degenerate tail probabilities}
\label{tab:SL-alpha-comparison}
\begin{tabular}{l|cc}
 & $\alpha\sim\mathrm{U}(0.01,0.025)$ & $\alpha=\alpha_0$ \\ \hline
Optimal threshold $\tau^*$ & $61087.34$ & $61233.08$ \\
Average first-best profit & $40675.64$ & $40675.64$ \\
Average optimal profit & $16831.00$ & $16922.08$ \\
Participation & $49.80\%$ & $49.92\%$ \\
Average information rent & $10327.74$ & $10163.67$
\end{tabular}
\end{table}

Finally, Table \ref{tab:SL-sensitivity} varies $\gamma$ in the heterogeneous-$\alpha$ case. A larger $\gamma$ both increases the curvature of $h_\gamma$ and raises $h_\gamma(x)$ pointwise for $x\in(0,1)$, thereby increasing the distorted valuation of risk. It consequently raises the deductible coefficient and the exclusion threshold while lowering participation and expected profit.
\begin{table}[H]
\centering
\caption{Sensitivity to the distortion parameter $\gamma$}
\label{tab:SL-sensitivity}
\begin{tabular}{c|ccccc}
$\gamma$ & $\theta_k^*/k$ & $\xi_k/k$ & $\tau^*$ & Average optimal profit & Participation \\ \hline
$0$    & $0.095310$ & $1.095310$ & $58721.26$ & $19586.57$ & $52.72\%$ \\
$0.25$ & $0.124062$ & $1.231348$ & $59885.63$ & $18188.71$ & $51.28\%$ \\
$0.50$ & $0.170941$ & $1.366310$ & $61087.34$ & $16831.00$ & $49.80\%$ \\
$0.75$ & $0.247597$ & $1.498996$ & $62316.75$ & $15526.15$ & $48.29\%$ \\
$1$    & $0.358836$ & $1.627174$ & $63552.31$ & $14295.60$ & $46.77\%$
\end{tabular}
\end{table}
\end{example}

\section{Class of quota-share policies}\label{sec:quotashare}

In this section, we consider the class of feasible menus from
$\mathcal{M}_2$, in which the only quota-share policies are considered.
For any such menu
\[M=\{v, I_{(a,k)}\mid v\in\mathcal{V}, I_{(a,k)}\in\mathcal{I}_2\text{ for all }(a,k)\in\bR_+\times K \}\in\mathcal{M}_2,\]
we have
$I_{(a,k)}(x)=\lambda_{(a,k)}x$ for all $(a,k)$, where the coinsurance rates $\lambda_{(a,k)}$ lie in the unit interval $[0,1]$. A each such policy is completely determined by the indirect utility function and the coinsurance rates, for notational simplicity, we write such menu as
$M=\{v, \lambda_{(a,k)}\}$.

In order for the menu $M=\{v, \lambda_{(a,k)}\}$ to be feasible,
the function $v\in\mathcal{V}$ and the numbers $\lambda_{(a,k)}\in[0,1]$ satisfy the IC$(a,k)$ constraint
\begin{equation}\label{vsub}
  v(a')-v(a)\geq \lambda_{(a,k)}(a'-a),\quad \text{for any }a'\in\mathbb{R}_+,
\end{equation}
for any $(a,k)\in\mathbb{R}_+\times K$. As $v$ is a convex function, the IC$(a,k)$ constraint (\ref{vsub}) simply states that $\lambda_{(a,k)}$ is a subgradient of $v$ at $a$. Therefore,
\[
\text{IC$(a,k)$}\Longleftrightarrow\begin{cases}
\lambda_{(a,k)}\in[v'_-(a),v'_+(a)]=\partial v(a),& \text{if }a>0,\\
\lambda_{(a,k)}\in[0,v'_+(0)],& \text{if }a=0,
\end{cases}
\]
where $\partial v(a)$ denotes the subdifferential of of the convex function $v$ at $a$.
If we set $v'_-(0)=0$ for consistency, then we arrive at the following characterization of the IC$(a,k)$ constraint  in terms of the subdifferential of $v$:
\begin{equation}\label{QS-IC}
\text{IC$(a,k)$}\Longleftrightarrow
\lambda_{(a,k)}\in \partial v(a).
\end{equation}

With the preparation above, we can now formally state the problem we aim to solve in this section: to maximize
\begin{equation}\label{Problem2}
J[M]=\int_{\cl{T}}\{\lambda_{(a,k)}(a - H[X_k])-v(a)\}\,{\mathbb{Q}}(da\times dk),
\end{equation}
over all menus $M=\{v, \lambda_{(a,k)}\}$ satisfying $v\in\cl{V}$ and $\lambda_{(a,k)}\in \partial v(a)$ for all $(a,k)\in\mathbb{R}_+\times K$.

 Let $P:=\{(a,k)\in\cl{T}: a\geq H[X_k]\}$ and $N:=\{(a,k)\in\cl{T}: a< H[X_k]\}$. For any feasible menu $M=\{v,\lambda_{(a,k)}\}\in\mathcal{M}_2$,
\begin{align*}
J[M]
&=\int_P \{\lambda_{(a,k)}(a - H[X_k])-v(a)\}\bQ(da\times dk)\\
&\qquad + \int_N \{\lambda_{(a,k)}(a - H[X_k])-v(a)\}\bQ(da\times dk),
\end{align*}
so for any fixed $v\in\cl{V}$ and subject to the IC${}_v(a,k)$ constraint $\lambda_{(a,k)}\in\partial v(a)$,
the intgrand is maximized
pointwise at
\begin{equation}\label{QS-lambda}
\lambda_{(a,k)}^*(v)=
\begin{cases}
v_+'(a), & \text{if } (a,k)\in P,\\
v_-'(a), & \text{if } (a,k)\in N.
\end{cases}
\end{equation}
 On $\{(a,k)\in\cl{T}: a= H[X_k]\}\subset P$, the choice of $\lambda_{(a,k)}\in \partial v(a)$ is indifferent, and for simplicity we choose $\lambda_{(a,k)}^*(v)=v'_+(a)$.

 With the optimal choice of the coinsurance rates $\lambda_{(a,k)}^*(v)$ as stated in (\ref{QS-lambda}) as functions of $v\in\cl{V}$, each feasible menu $M=\{v,\lambda_{(a,k)}\}\in\mathcal{M}_2$ is essentially represented by  $v\in\mathcal{V}$ alone, and Problem (\ref{Problem2}) is now reduced to maximizing
\begin{align}
J[v]&:=\int_P \{v_+'(a)(a - H[X_k])-v(a)\}\bQ(da\times dk)\notag\\
&\qquad + \int_N \{v_-'(a)(a - H[X_k])-v(a)\}\bQ(da\times dk)\label{QS-J}
\end{align}
over $v\in\mathcal{V}$.

For any $t\in\overline{\mathbb{R}}_+$, define $\phi_t\in\mathcal{V}$ by $\phi_t(\cdot)=(\cdot-t)_+$. The next result expresses the objective function $J[v]$ in (\ref{QS-J}) as a weighted average of
$J[\phi_t]$, which then allows us to further reduce Problem (\ref{QS-J}), which is infinite-dimensional, into a one-dimensional problem. To this end, we recall the following special case of the useful Breeden-Litzenberger formula, which expresses every increasing convex function $g:\mathbb{R}_+\to\mathbb{R}$ as
\[g(a)=g(0)+\int_{[0,\infty)} (a-t)_+\,\gamma_g(dt), \quad a\geq 0,\]
where $\gamma_g$ is a Radon measure on $\mathbb{R}_+$ defined via $\gamma_g[0,x]=g_+'(x)$ for $x\geq 0$. More discussion can be found in \cite{follmer2011stochastic}.

\begin{proposition}\label{Jv}
For any $v\in\mathcal{V}$,  define the Radon measure $\gamma_v$ on $\mathbb{R}_+$ by $\gamma_v[0,a]=v'_+(a)$, $a\geq 0$. Then
\[J[v]=\int_{\mathbb{R}_+}J[\phi_t]\gamma_v(dt).\]
\end{proposition}
{\em Proof. } Fix $v\in\mathcal{V}$, and define the measure $\gamma_v$ on $\mathbb{R}_+$ as in the statement of the theorem. By the Breeden-Litzenberger formula, we have
\[v(a)=\int_{[0,\infty)} (a-t)_+\,\gamma_v(dt), \quad a\geq 0.\]
On the other hand, we can also express $v'_+(a)$ as
\[v'_+(a)=\int_{[0,\infty)} \mathbf{1}_{\{t\leq a\}}\,\gamma_v(dt), \quad a\geq 0.\]
Putting these two expressions in (\ref{QS-J}),  and noting $\gamma_v\{a\}=v_+'(a)-v'_-(a)$, we get
\begin{align*}
    J[v]&=\int_P \{v_+'(a)(a - H[X_k])-v(a)\}\bQ(da\times dk)\\
&\qquad + \int_N \{v_-'(a)(a - H[X_k])-v(a)\}\bQ(da\times dk)\\
    &=\int_{\cl{T}}\left[\int_{[0,\infty)}\{\mathbf{1}_{\{t\leq a\}}(a - H[X_k])-(a-t)_+\}\,\gamma_v(dt)\right]\bQ(da\times dk)\\
    &\qquad -\int_N\{(v_+'(a)-v'_-(a))(a - H[X_k])\}\bQ(da\times dk)\\
    &=\int_{[0,\infty)}\left[\int_{\cl{T}}\{\mathbf{1}_{\{t\leq a\}}(a - H[X_k])-(a-t)_+\}\,\bQ(da\times dk)\right]\gamma_v(dt)\\
    &\qquad -\int_N\{\gamma_v\{a\}(a - H[X_k])\}\bQ(da\times dk).
\end{align*}
On the other hand, for $t\geq 0$, we have also
\begin{align}
    J[\phi_t]&=\int_P \{(\phi_t)_+'(a)(a - H[X_k])-\phi_t(a)\}\bQ(da\times dk)\notag \\
&\qquad + \int_N \{(\phi_t)_-'(a)(a - H[X_k])-\phi_t(a)\}\bQ(da\times dk)\notag \\
    &=\int_{\cl{T}} \{\mathbf{1}_{\{t\leq a\}}(a - H[X_k])-(a-t)_+\}\bQ(da\times dk)\notag \\
&\qquad - \int_N \{\mathbf{1}_{\{t=a\}}(a - H[X_k])\}\bQ(da\times dk).\label{ForEg2}
\end{align}
Putting the above two equations together, we have
\begin{align*}
    J[v] &= \int_{[0,\infty)}J[\phi_t]\,\gamma_v(dt)+\int_{[0,\infty)}\left\{\int_N \{\mathbf{1}_{\{t=a\}}(a - H[X_k])\}\,\bQ(da\times dk)\right\}\gamma_v(dt)\\
    &\qquad -\int_N\{\gamma_v\{a\}(a - H[X_k])\}\bQ(da\times dk).
\end{align*}
The second and the third terms on the right are indeed the same:
\begin{align*}
&\int_{[0,\infty)}\left\{\int_N \{\mathbf{1}_{\{t=a\}}(a - H[X_k])\}\,\bQ(da\times dk)\right\}\gamma_v(dt)\\
&\qquad =\int_N\left\{\int_{[0,\infty)} \{\mathbf{1}_{\{t=a\}}(a - H[X_k])\}\,\gamma_v(dt)\right\}\bQ(da\times dk)   \\
    &\qquad =\int_N\left\{\gamma_v\{a\}(a - H[X_k])\right\}\bQ(da\times dk),
\end{align*}
and hence we obtain
\[J[v]= \int_{[0,\infty)}J[\phi_t]\,\gamma_v(dt),\]
as desired.\hfill $\square$


As a consequence of the previous theorem, we can prove that the objective function $J[v]$ in (\ref{QS-J}) attains its maximum over $\mathcal{V}$ at $\phi_{\tau^*}$ for some $\tau^*\in\overline{\mathbb{R}}_+$.

\begin{lemma}\label{L:QS-existence}
Under Condition (\ref{Integrability}), the map $t\mapsto J[\phi_t]$ is upper semicontinuous on $\overline{\mathbb{R}}_+$, endowed with its one-point compactification topology, where $\phi_\infty\equiv0$ and $J[\phi_\infty]=0$. Consequently, $J[\phi_t]$ attains its maximum on $\overline{\mathbb{R}}_+$.
\end{lemma}
{\em Proof.} For every finite $t\geq0$, equation (\ref{ForEg2}) can be equivalently written as
\[
J[\phi_t]=\int_{\mathcal{T}}\psi_t(a,k)\,\mathbb{Q}(da\times dk),
\]
where
\[
\psi_t(a,k):=(t-H[X_k])\mathbf{1}_{\{a>t\}}+(t-H[X_k])_+\mathbf{1}_{\{a=t\}}.
\]
For every fixed $(a,k)$, the map $t\mapsto\psi_t(a,k)$ is upper semicontinuous on $\overline{\mathbb{R}}_+$ and satisfies
\[
-H[X_k]\leq\psi_t(a,k)\leq a.
\]
Condition (\ref{Integrability}) and the reverse Fatou lemma therefore imply that $t\mapsto J[\phi_t]$ is upper semicontinuous. Since $\overline{\mathbb{R}}_+$ is compact, the maximum is attained.\hfill $\square$

\begin{theorem}\label{QS-Op}
Let
\begin{equation}\label{QS-Tau*}
\tau^*\in\operatorname*{arg\,max}_{t\in\overline{\mathbb{R}}_+}J[\phi_t].
\end{equation}
Then
\[\max_{v\in\cl{V}} J[v]=J[\phi_{\tau^*}].\]
\end{theorem}
{\em Proof. } By Proposition \ref{Jv}, for any $v\in\mathcal{V}$, we can express $J[v]$
as
\[J[v]= \int_{[0,\infty)}J[\phi_t]\,\gamma_v(dt),\]
where $\gamma_v$ is the measure on $\mathbb{R}_+$ defined via $\gamma_v[0,a]=v'_+(a)$ for $a\geq 0$. {\color{Brown}Moreover, $J[\phi_{\tau^*}]\geq J[\phi_\infty]=0$.} If $\tau^*$ is defined as in (\ref{QS-Tau*}), then $J[\phi_t]\leq J[\phi_{\tau^*}]$ for every $t\geq 0$, and hence
\begin{align*}
    J[v]    \leq\int_{[0,\infty)}J[\phi_{\tau^*}]\,\gamma(dt)=J[\phi_{\tau^*}]\gamma_v[0,\infty)  =J[\phi_{\tau^*}]v'_+(+\infty)
    \leq J[\phi_{\tau^*}],
\end{align*}
where the last inequality follows from Lemma \ref{L:IndirectUtility}. We conclude that every $v\in\mathcal{V}$ is dominated by $\phi_{\tau^*}$, and hence the optimality of $\phi_{\tau^*}$ follows.\hfill $\square$

After identifying that the optimal indirect utility function  is $\phi_{\tau^*}$, we can then derive the corresponding optimal coinsurance rates and premiums. They are presented in the next theorem, which is the main result of this secton:

\begin{theorem}\label{T_QS-Op}
For Problem (\ref{Problem2}), the optimal coinsurance rate  $\lambda_{(a,k)}^*$ and the premium $P^*_{(a,k)}$ corresponding to the quota-share policy with such coinsurance rate are given respectively by
\[
\lambda_{(a,k)}^* =
\begin{cases}
1 & \text{if }a>\tau^* \\
1 & \text{if }a=\tau^*\geq H[X_k] \\
0 & \text{if }a=\tau^*< H[X_k] \\
0 & \text{if }a<\tau^*, \\
\end{cases}\quad
\text{and}\quad
P_{(a,k)}^* =
\begin{cases}
\tau^* & \text{if }a>\tau^* \\
\tau^* & \text{if }a=\tau^*\geq H[X_k] \\
0 & \text{if }a=\tau^*< H[X_k] \\
0 & \text{if }a<\tau^*.
\end{cases}
\]
\end{theorem}
{\em Proof. } From Theorem \ref{QS-Op}, the optimal indirect utility function $v_{M^*}$ associated with the optimal menu $M^*$ is given by $\phi_{\tau^*}$, where $\tau^*$ is defined in (\ref{QS-Tau*}). The optimal coinsurance rate $\lambda_{(a,k)}^*$ can then by obtained from (\ref{QS-lambda}), so that $\lambda_{(a,k)}^*=\lambda_{(a,k)}^*(\phi_{\tau^*})$. By Lemma \ref{L:IndirectUtility},
\[P_{(a,k)}^*=I_{(a,k)}^*(a)-v_{M^*}(a)=\lambda_{(a,k)}^*a-(a-\tau^*)_+,\]
from which we can obtain the expression of the corresponding premium $P_{(a,k)}^*$ as stated in the theorem.\hfill $\square$

The following example illustrates Theorem \ref{T_QS-Op} and the role of heterogeneity in the tail probability.
\begin{example}\label{EG:QS}
We adopt the same numerical setting of Example \ref{EG:ST}. Since
\[
H[X_k]=1.1k\left(1+\frac{\gamma}{2}\right)
\leq1.65k<\underline c k\leq a,
\qquad \underline c=-\log(0.025),
\]
$N=\varnothing$, and (\ref{ForEg2}) reduces to
\[
\max_{t\in\overline{\mathbb{R}}_+}\frac{1}{20000}
\int_{5000}^{25000}
\left[t-1.1\left(1+\frac{\gamma}{2}\right)k\right]
\Pr(a>t\mid k)\,dk.
\]
For the baseline $\gamma=0.5$, the optimal threshold, indirect utility, and average profit are
\[
\tau^*_{\rm QS}\approx61166.37,\qquad
v^*_{\rm QS}(a)=(a-61166.37)_+,\qquad
J[v^*_{\rm QS}]\approx16744.61.
\]
Hence
\[
(\lambda_{(a,k)}^*,P_{(a,k)}^*)=
\begin{cases}
(1,61166.37), & a\geq61166.37,\\
(0,0), & a<61166.37.
\end{cases}
\]

For comparison, remove heterogeneity in $\alpha$ while preserving mean willingness to pay by setting
$\alpha_0=\exp(\mathbb{E}[\log\alpha])\approx0.016941$. Then $a=c_0k$, where $c_0=-\log\alpha_0\approx4.078019$, and
\[
\max_{5000c_0\leq t\leq25000c_0}\frac{1}{20000}
\int_{t/c_0}^{25000}(t-1.375k)\,dk,
\qquad
\tau^*_{\rm QS,0}=\frac{25000c_0^2}{2c_0-1.375}\approx61311.55.
\]
Equivalently, full reinsurance is offered at the common premium $61311.55$ precisely when
$k\geq k^*_{\rm QS,0}=\tau^*_{\rm QS,0}/c_0\approx15034.64$.

\begin{table}[H]
\centering
\caption{Quota-share outcomes under heterogeneous and degenerate tail probabilities}
\label{tab:QS-alpha-comparison}
\begin{tabular}{l|cc}
 & $\alpha\sim\mathrm{U}(0.01,0.025)$ & $\alpha=\alpha_0$ \\ \hline
Optimal threshold & $61166.37$ & $61311.55$ \\
Average first-best profit & $40545.28$ & $40545.28$ \\
Average optimal profit & $16744.61$ & $16835.35$ \\
Participation & $49.71\%$ & $49.83\%$ \\
Average information rent & $10288.42$ & $10124.54$
\end{tabular}
\end{table}

Under full information all types are served, whereas asymmetric information excludes approximately half of the market. Holding mean willingness to pay fixed, heterogeneity in $\alpha$ raises information rent by about $1.6\%$ and lowers profit by about $0.5\%$, but has little effect on aggregate participation; its main effect is to change which intermediate-$k$ agents participate. In both cases all participating types are pooled at the same full-reinsurance contract, unlike the $k$-dependent deductibles and premiums in Example \ref{EG:ST}.
\end{example}


\section{Class of change-loss policies}\label{sec:changeloss}

In this section, we consider the class of feasible menus from
$\mathcal{M}_3$.
For any such menu
\[M=\{v, I_{(a,k)}\mid v\in\mathcal{V}, I_{(a,k)}\in\mathcal{I}_3\text{ for all }(a,k)\in\bR_+\times K \}\in\mathcal{M}_3,\]
we have
$I_{(a,k)}(x)=\lambda_{(a,k)}(x-d_{(a,k)})_+$ for all $(a,k)$, where
\[\lambda_{(a,k)}\in[0,1]\quad\text{and}\quad d_{(a,k)} \in\overline{\mathbb{R}}_+\]
are the coinsurance rate and the deductible respectively.
The null policy $I(x)\equiv 0$ can be represented by taking either $d=+\infty$ or $\lambda=0$. For notational simplicity, we write such menu as
$M=\{v, \lambda_{(a,k)},d_{(a,k)}\}$.

In order for the menu $M=\{v, \lambda_{(a,k)},d_{(a,k)}\}$ to be feasible,
the indirect utility function $v\in\mathcal{V}$,  the coinsurance rates $\lambda_{(a,k)}\in[0,1]$ and the deductibles $d_{(a,k)}\in\overline{\mathbb{R}}_+$ must satisfy the IC$(a,k)$ constraint
\[
v(a')-v(a)\geq \lambda_{(a,k)}(a'-d_{(a,k)})_+-\lambda_{(a,k)}(a-d_{(a,k)})_+,\quad \text{for any }a'\in\mathbb{R}_+,
\]
for any $(a,k)\in\mathbb{R}_+\times K$. The problem we aim to solve in this section is to maximize
\begin{equation}\label{Problem3}
J[M]=\int\{\lambda_{(a,k)}[(a-d_{(a,k)})_+ - H(X_k-d_{(a,k)})_+]-v(a)\}\bQ(da\times dk),
\end{equation}
over all feasible menus $M\in\mathcal{M}_3$.

Our first key step in solving this problem is to obtain necessary conditions that any feasible menu in $\cl{M}_3$ must satisfy.

\begin{lemma}\label{CL-v}
For any feasible menu $M=\{v,\lambda_{(a,k)}, d_{(a,k)}\}\in\mathcal{M}_3$, it must hold that
\begin{equation}\label{CL-IC1}
\begin{cases}
v_-'(a)\leq \lambda_{(a,k)} \mathbf{1}_{\{a>d_{(a,k)}\}}\leq \lambda_{(a,k)} \mathbf{1}_{\{a\geq d_{(a,k)}\}}\leq v_+'(a),& a>0,\\
\lambda_{(0,k)} \mathbf{1}_{\{0\geq d_{(0,k)}\}}\leq v_+'(0),& a=0,
\end{cases}
\end{equation}
for any $k\in K$.
\end{lemma}
{\em Proof.} Fix any $a>0$ and $k\in K$.  Let $g(x)=\lambda_{(a,k)}(x-d_{(a,k)})_+$, $x\geq 0$. The right-hand derivative is $g_+'(x)=\lambda_{(a,k)} \mathbf{1}_{\{x\geq d_{(a,k)}\}}$. Now consider $a'\downarrow a$. By the IC$(a,k)$ constraint,
\[ v_+'(a)=\lim_{a'\downarrow a}\frac{v(a')-v(a)}{a'-a}\geq \lim_{a'\downarrow a}\frac{g(a')-g(a)}{a'-a}=g_+'(a)=\lambda_{(a,k)} \mathbf{1}_{\{a\geq d_{(a,k)}\}}.\]
Similarly, when $a>0$, we can obtain
\[0\leq v_-'(a)\leq g_-'(a)=\lambda_{(a,k)} \mathbf{1}_{\{a>d_{(a,k)}\}}\leq \lambda_{(a,k)} \mathbf{1}_{\{a\geq d_{(a,k)}\}}\leq v_+'(a).\]
The case for $a=0$ is similar and is omitted.\hfill $\square$


With Lemma \ref{CL-v} on hand, we can considerably simplify the IC constraints. To this end, for any fixed $v\in\cl{V}$, we define \[\tau_v:=\sup\{a: v(a)=0\}\in\overline{\mathbb{R}}_+.\]
It is possible that $\tau_v=0$.

\begin{lemma}\label{L:CL-IC}
Let $v\in\cl{V}$ be fixed. For any $(a,k)\in\mathbb{R}_+\times K$,

\noindent (i) if $a\ne\tau_v$,
\begin{equation}\label{CL-IC2a}
\mathrm{IC}_v(a,k)\Longleftrightarrow
\begin{cases}
\lambda_{(a,k)}=0\text{ or } v(a')\geq\lambda_{(a,k)}(a'-d_{(a,k)})_+\quad\forall a', & \text{if } a<\tau_v,\\
 \lambda_{(a,k)}\in\partial v(a), d_{(a,k)}\leq a-v(a)/\lambda_{(a,k)}, & \text{if } a>\tau_v;
\end{cases}
\end{equation}
(ii) if $a=\tau_v$,
\begin{equation}\label{CL-IC2b}
\mathrm{IC}_v(\tau_v,k)\Longleftrightarrow
\begin{cases}
\lambda_{(\tau_v,k)}=0\,\text{ or } \,v(a')\geq\lambda_{(\tau_v,k)}(a'-d_{(\tau_v,k)})_+ \quad\forall a', & \text{if } v_+'(\tau_v)=0,\\
\begin{cases}
d_{(\tau_v,k)}\leq\tau_v,\lambda_{(\tau_v,k)}\in\partial v(\tau_v)\text{ or}, \\
d_{(\tau_v,k)}>\tau_v,v(a')\geq\lambda_{(\tau_v,k)}(a'-d_{(\tau_v,k)})_+\quad\forall a' ,
\end{cases}& \text{if } v_+'(\tau_v)>0.
\end{cases}
\end{equation}
\end{lemma}
{\em Proof. } (i) First we consider the case when $a<\tau_v$. Assume that the IC${}_v(a,k)$ constraint holds. From (\ref{CL-IC1}), $\lambda_{(a,k)} \mathbf{1}_{\{a>d_{(a,k)}\}}= \lambda_{(a,k)} \mathbf{1}_{\{a\geq d_{(a,k)}\}}=0$, so either (i) $\lambda_{(a,k)}=0$, or (ii) $\lambda_{(a,k)}>0$, $d_{(a,k)}>a$. In case (ii), the IC${}_v(a,k)$ constraint reduces to that $v(a')\geq\lambda_{(a,k)}(a'-d_{(a,k)})_+$ for all $a'$. Conversely, the IC${}_v(a,k)$ constraint is trivially satisfied if $\lambda_{(a,k)}=0$; moreover, if $v(a')\geq\lambda_{(a,k)}(a'-d_{(a,k)})_+$ for all $a'$, then
\[v(a')-v(a)=v(a')\geq\lambda_{(a,k)}(a'-d_{(a,k)})_+\geq\lambda_{(a,k)}(a'-d_{(a,k)})_+-\lambda_{(a,k)}(a-d_{(a,k)})_+\]
for all $a'$, and hence the IC${}_v(a,k)$ constraint also holds true.

Next we consider the case when $a>\tau_v$. Assume that the IC${}_v(a,k)$ constraint holds. From (\ref{CL-IC1}), $0<v_-'(a)\leq\lambda_{(a,k)} \mathbf{1}_{\{a>d_{(a,k)}\}}\leq \lambda_{(a,k)} \mathbf{1}_{\{a\geq d_{(a,k)}\}}\leq v_+'(a)$. It follows that $d_{(a,k)}<a$ and $\lambda_{(a,k)}\in\partial v(a)$. Putting $a'=0$ in the IC${}_v(a,k)$ constraint gives
\[v(0)-v(a)\geq\lambda_{(a,k)}(0-d_{(a,k)})_+-\lambda_{(a,k)}(a-d_{(a,k)})_+=-\lambda_{(a,k)}(a-d_{(a,k)}).\]
Rearranging this inequality yields
\[d_{(a,k)}\leq a-\frac{v(a)}{\lambda_{(a,k)}}.\]
Conversely, assume that $\lambda_{(a,k)}\in\partial v(a)$ and $ d_{(a,k)}\leq a-v(a)/\lambda_{(a,k)}$. We verify that the IC${}_v(a,k)$ constraint holds true by considering $a'\geq d_{(a,k)}$ and $a'< d_{(a,k)}$ separately. For $a'\geq d_{(a,k)}$, since $\lambda_{(a,k)}\in\partial v(a)$, it follows that
\begin{align*}
    v(a')-v(a) &\geq \lambda_{(a,k)}(a'-a)\\
    &= \lambda_{(a,k)}[(a'-d_{(a,k)})-(a-d_{(a,k)})]\\
    &\geq \lambda_{(a,k)}[(a'-d_{(a,k)})_+-(a-d_{(a,k)})_+],
\end{align*}
which is just the desired IC${}_v(a,k)$ constraint. For $a'< d_{(a,k)}$, we have
\begin{align*}
    v(a')-v(a)&\geq -v(a)\\
    &\geq -\lambda_{(a,k)}(a-d_{(a,k)})\\
    &= -\lambda_{(a,k)}(a-d_{(a,k)})_+\\
    &= \lambda_{(a,k)}(a'-d_{(a,k)})_+-\lambda_{(a,k)}(a-d_{(a,k)})_+,
\end{align*}
as desired.

(ii) For $a=\tau_v$, the case of $v_+'(\tau_v)=0$ is identical to that when $a<\tau_v$.

The only case that remains is $a=\tau_v$ with $v_+'(\tau_v)>0$. Assume that the IC${}_v(\tau_v,k)$ constraint holds true. If $d_{(\tau_v,k)}\leq\tau_v$, the IC${}_v(\tau_v,k)$ constraint becomes
\[ v(a')\geq \lambda_{(\tau_v,k)}(a'-d_{(\tau_v,k)})_+-\lambda_{(\tau_v,k)}(\tau_v-d_{(\tau_v,k)}),\quad\forall a',\]
which is always true when $a'\leq d_{(\tau_v,k)}$, and hence it is equivalent to
\[ v(a')\geq \lambda_{(\tau_v,k)}(a'-d_{(\tau_v,k)})-\lambda_{(\tau_v,k)}(\tau_v-d_{(\tau_v,k)})=\lambda_{(\tau_v,k)}(a'-\tau_v),\quad\forall a'> d_{(\tau_v,k)},\]
which is further seen to be equivalent to $\lambda_{(\tau_v,k)}\in\partial v(\tau_v)$ as $v$ is convex. On the other hand, if $d_{(\tau_v,k)}>\tau_v$, the IC${}_v(\tau_v,k)$ constraint becomes
\[ v(a')\geq \lambda_{(\tau_v,k)}(a'-d_{(\tau_v,k)})_+,\quad\forall a'.\]
This completes the proof.\hfill $\square$

At the moment, we still do not yet know the optimal indirect utility function $v^*$ and its corresponding $\tau_{v^*}$. Nevertheless, by making use of the previous lemma, we can show that $v\in\cl{V}$ with $\tau_v<L$ cannot be optimal for Problem (\ref{Problem3}), where
\[L:= \inf\{a\in\bR_+: \exists k\in K \text{ such that }(a,k)\in\cl{T}\}.\]
This fact will prove to be crucial in simplifying the analysis of Problem (\ref{Problem3}).

\begin{lemma}\label{L:tau<L}
Every $v\in\cl{V}$ with $\tau_v<L$ is not optimal for Problem (\ref{Problem3}).
\end{lemma}
{\em Proof. } We only need to consider the case where $L>0$. Suppose that $\tau_v<L$.  Pick any $\tilde{\tau}\in(\tau_v,L)$, and define
\[\tilde{v}(a) = (v(a)-v(\tilde{\tau}))_+,\quad a\in\bR_+.\]
It is evident that $\tilde{v}\in\cl{V}$ with $\tau_v<\tau_{\tilde{v}}=\tilde{\tau}<L$,  $\partial v(a)=\partial\tilde{v}(a)$ and $\tilde{v}(a)<v(a)$ for all $a> \tilde{\tau}$. For any $(a,k)\in\cl{T}$, it then holds that $a\geq L>\tau_{\tilde{v}}>\tau_v$.

Now consider an arbitrary feasible $M=\{v,\lambda_{(a,k)},d_{(a,k)}\}\in\cl{M}_3$. For any $(a,k)\in\cl{T}$,
 $\lambda_{(a,k)}$ and $d_{(a,k)}$ satisfy the IC${}_v(a,k)$ constraint (\ref{CL-IC2a}) in Lemma \ref{L:CL-IC}:
\[ \lambda_{(a,k)}\in\partial v(a),\quad\text{and}\quad d_{(a,k)}\leq a-v(a)/\lambda_{(a,k)}.\]
If we replace $v$ by $\tilde{v}$ and leave $\lambda_{(a,k)}$ and $d_{(a,k)}$ unchanged, we obtain a new menu $\tilde{M}=\{\tilde{v},\lambda_{(a,k)},d_{(a,k)}\}$. This  is also a feasible menu because for any $(a,k)\in\cl{T}$,
\[ \lambda_{(a,k)}\in\partial v(a)=\partial\tilde{v}(a),\quad\text{and}\quad d_{(a,k)}\leq a-v(a)/\lambda_{(a,k)}<a-\tilde{v}(a)/\lambda_{(a,k)},\]
that is, $\lambda_{(a,k)}$ and $d_{(a,k)}$  satisfy the IC${}_{\tilde{v}}(a,k)$ constraint. Since $\tilde{v}(a)<v(a)$ for all $a\geq L$, it is then clear from the objective function  in (\ref{Problem3}) that $J[v]<J[\tilde{v}]$, and hence $v$ cannot be optimal.\hfill $\square$


Thanks to Lemma \ref{L:tau<L}, we can narrow down our search of the optimal indirect utility for Problem (\ref{Problem3}) to the following subset of $\cl{V}$:
\[\cl{V}_0=\{v\in\cl{V}: \tau_v\geq L\}.\]

From now on, we impose the following assumption:
\begin{equation}\label{OnlyAssumption}
    \sup_{k\in K}\theta_k^* \leq L:= \inf\{a\in\bR_+: \exists k\in K \text{ such that }(a,k)\in\cl{T}\}\in(0,\infty).
\end{equation}
While Assumption (\ref{OnlyAssumption}) might  not be  interpreted directly, it is nevertheless implied by the following stronger condition:
\begin{equation}\label{OnlyAssumption**}
\sup_{k\in K}H[X_k] \leq L.
\end{equation}
Indeed, the optimality of $\theta_k^*$ implies $\xi_k\leq H[X_k]$, while $\theta_k^*\leq \xi_k$. Hence, (\ref{OnlyAssumption**}) immediately implies (\ref{OnlyAssumption}). Condition (\ref{OnlyAssumption**}) also admits a natural pointwise-surplus interpretation. Under full reinsurance, the IR constraint implies that an insurer of type $(a,k)$ is willing to pay at most $a$, whereas the reinsurer evaluates the indemnity at the risk-adjusted cost $H[X_k]$. Thus, full reinsurance generates a nonnegative pointwise surplus whenever $a\geq H[X_k]$. Since every feasible type satisfies $a\geq L$, Condition (\ref{OnlyAssumption**}) guarantees this inequality for all $(a,k)\in\cl{T}$.


The strategy to solve Problem (\ref{Problem3}) is the same as before: we first fix $v\in \cl{V}_0$ and maximize the integrand of the objective function pointwise over $d_{(a,k)}$ and $\lambda_{(a,k)}$, subject to the IC${}_{v}(a,k)$ constraints (\ref{CL-IC2a}) and (\ref{CL-IC2b}) in Lemma \ref{L:CL-IC}, to obtain $d^*_{(a,k)}$ and $\lambda^*_{(a,k)}$ as functions of $v$, then we optimize over the choice of $v\in\cl{V}_0$. To this end, we denote by $\cl{T}_+$ the subset of agents $(a,k)\in\cl{T}$ with $a\geq\xi_k$, and $\cl{T}_-$ the subset of agents $(a,k)\in\cl{T}$ with $a<\xi_k$:
\[\cl{T}_+:=\{(a,k)\in\cl{T}: a\geq\xi_k\},\quad\text{and}\quad\cl{T}_-:=\{(a,k)\in\cl{T}: a<\xi_k\}.\]
The next results identifies the optimal coinsurance rates and deductibles when the indirect utility function is fixed.

\begin{proposition}\label{CL-Op}
Fix any $v\in\cl{V}_0$. For any $(a,k)\in \cl{T}$, the optimal coinsurance rates  and the optimal deductibles  for Problem (\ref{Problem3}), both as functions of $v$,  are given respectively by
\[\lambda_{(a,k)}^*(v) =
\begin{cases}
v_+'(a) & \text{if }a\geq\tau_v, (a,k)\in\cl{T}_+ \\
v_-'(a) & \text{if }a\geq\tau_v, (a,k)\in\cl{T}_- \\
0 & \text{if }a<\tau_v,
\end{cases}
\qquad
d_{(a,k)}^*(v) =
\begin{cases}
\theta_k^* & \text{if }a\geq\tau_v\\
\text{irrelevant} & \text{if }a<\tau_v,
\end{cases}\]
\end{proposition}
{\em Proof. } For any $(a,k)\in \cl{T}$, maximizing the integrand of the objective function of Problem (\ref{Problem3}) is the same as maximizing
\begin{equation}\label{CL-integrand}
F_{(a,k)}(\lambda_{(a,k)},d_{(a,k)}):=\lambda_{(a,k)}[(a-d_{(a,k)})_+ - H(X_k-d_{(a,k)})_+]
\end{equation}
over $d_{(a,k)}$ and $\lambda_{(a,k)}$, subject to the IC${}_{v}(a,k)$ constraints (\ref{CL-IC2a}) and (\ref{CL-IC2b}), which is done according the following cases:

\underline{Case (i):} Suppose that $a<\tau_v$. We want to maximize $F_{(a,k)}(\lambda_{(a,k)},d_{(a,k)})$ subject to either
    \[\text{(a) }\lambda_{(a,k)}=0 \quad\text{ or }\quad\text{(b) } \lambda_{(a,k)}>0, v(a')\geq \lambda_{(a,k)}(a'-d_{(a,k)})_+,\quad a'\in\bR_+.\]
For (a), $F(\lambda_{(a,k)},d_{(a,k)})$ equals 0 no matter what $d_{(a,k)}$ is. For (b), putting $a'=\tau_v$ in the constraint gives
\[0=v(\tau_v)\geq \lambda_{(a,k)}(\tau_v-d_{(a,k)})_+,\]
and hence
$d_{(a,k)}\geq\tau_v$. Therefore,
\[F_{(a,k)}(\lambda_{(a,k)},d_{(a,k)})=-\lambda_{(a,k)} H(X_k-d_{(a,k)})_+<0\]
because $d_{(a,k)}\geq \tau_v>a$. Obviously, (1) is better, and so $\lambda^*_{(a,k)}(v)=0$. The resulting policy is a null policy, and the value of the deductible is irrelevant.

\underline{Case (ii):} Suppose that $a>\tau_v$. To maximize $F_{(a,k)}(\lambda_{(a,k)},d_{(a,k)})$  subject to the IC${}_{v}(a,k)$ constraint
        \[ \lambda_{(a,k)}\in\partial v(a),\quad\text{and}\quad d_{(a,k)}\leq a-v(a)/\lambda_{(a,k)},\]
we first fix any $\lambda\in\partial v(a)$, and obtain the corresponding $d^*_{(a,k)}(\lambda)$ by maximizing
\begin{equation}\label{CL_max}
F_{(a,k)}(\lambda,d)=\lambda\left\{(a-d)_+ - H(X_k-d)_+\right\},
\end{equation}
over $d\in[0,a-v(a)/\lambda]$. This gives
\[d^*_{(a,k)}(\lambda)=\theta^*_k\wedge(a-v(a)/\lambda).\]
On the other hand, the convexity of $v$ and (\ref{OnlyAssumption}) imply that
\[a-v(a)/\lambda\geq \tau_v\geq L\geq\theta_k^*.\]
This leads to $d^*_{(a,k)}(\lambda)=\theta^*_k$.
The value of $\lambda^*_{(a,k)}$ can now obtained by solving
\[
\max_{\lambda\in\partial v(a)}F_{(a,k)}(\lambda,d^*_{(a,k)}(\lambda)).
\]
It follows from (\ref{OnlyAssumption})  that for any $(a,k)\in\cl{T}$,
\begin{align*}
F_{(a,k)}(\lambda,d^*_{(a,k)}(\lambda))&=\lambda\left\{(a-\theta^*_k)_+ - H(X_k-\theta^*_k)_+\right\}\\
&=\lambda\left\{a-\theta^*_k- H(X_k-\theta^*_k)_+\right\}\\
&=\lambda(a-\xi_k).
\end{align*}
The coefficient of $\lambda$, $a-\xi_k$, is positive when $(a,k)\in\cl{T}_+$ and negative when $(a,k)\in\cl{T}_-$.
Therefore, $\lambda_{(a,k)}^*(v)=v_+'(a)$ when $(a,k)\in\cl{T}_+$ and $\lambda_{(a,k)}^*(v)=v_-'(a)$ when $(a,k)\in\cl{T}_-$. As $d^*_{(a,k)}(\lambda)=\theta^*_k$ is independent of $\lambda$, $d^*_{(a,k)}(v)=d^*_{(a,k)}(\lambda^*)=\theta^*_k$.

\underline{Case (iii):} Finally, we consider the case when $a=\tau_v$. There are two subcases to deal with:

Subcase (1): $v_+'(\tau_v)=0$. This subcase is similar to the case when $a<\tau_v$: under the IC${}_v(\tau_v,k)$ constraint, the integrand $F_{(\tau_v,k)}(\lambda_{(\tau_v,k)},d_{(\tau_v,k)})$ in (\ref{Problem3})   is maximized at $\lambda^*_{(\tau_v,k)}(v)=0=v_-'(\tau_v)=v_+'(\tau_v)$ and the value of $d^*_{(\tau_v,k)}(v)$ can be set arbitrarily. To be consistent with the subcase below, we simply set $d^*_{(\tau_v,k)}(v)=\theta_k^*$.

Subcase (2): $v_+'(\tau_v)>0$. The IC${}_v(\tau_v,k)$ constraint states that
either (a) $d_{(\tau_v,k)}>\tau_v$ and $v(a')\geq\lambda_{(\tau_v,k)}(a'-d_{(\tau_v,k)})_+$ for all $a'$, or (b) $d_{(\tau_v,k)}\leq \tau_v$ and $\lambda_{(\tau_v,k)}\in\partial v(\tau_v)=[0,v'_+(\tau_v)]$. For scenario (a),
\[F_{(\tau_v,k)}(\lambda_{(\tau_v,k)},d_{(\tau_v,k)})=-\lambda_{(\tau_v,k)}H(X_k-d_{(\tau_v,k)})_+\leq 0,\]
whose maximum value is 0 attained at $\lambda^*_{(\tau_v,k)}=0$. For scenario (b), \[F_{(\tau_v,k)}(\lambda_{(\tau_v,k)},d_{(\tau_v,k)})=\lambda_{(\tau_v,k)}[\tau_v-d_{(\tau_v,k)} - H(X_k-d_{(\tau_v,k)})_+]\]
is maximized at $d^*_{(\tau_v,k)}=\theta_k^*\wedge\tau_v=\theta_k^*$;
moreover, as
\[F_{(\tau_v,k)}(\lambda_{(\tau_v,k)},\theta_k^*)=\lambda_{(\tau_v,k)}[\tau_v-\theta_k^* - H(X_k-\theta_k^*)_+]=\lambda_{(\tau_v,k)}(\tau_v-\xi_k)\]
we see that
$\lambda^*_{(\tau_v,k)}=v_+'(\tau_v)$ if $(\tau_v,k)\in\cl{T}_+$ and $\lambda^*_{(\tau_v,k)}=0$ if $(\tau_v,k)\in\cl{T}_-$. As scenario (b) is always as good as scenario (a), we take $\lambda^*_{(\tau_v,k)}(v)$ and $d^*_{(\tau_v,k)}(v)$ from scenario (b).\hfill$\square$

With the optimal choice of $\lambda_{(a,k)}^*(v)$ and  $d_{(a,k)}^*(v)$ as functions of $v$ as stated in Proposition \ref{CL-Op},
we can restrict to those feasible menus of the form $M=\{v,\lambda_{(a,k)}(v),d_{(a,k)}(v)\}\in\mathcal{M}_3$, which can  be  represented by  $v\in\mathcal{V}_0$ alone. For these feasible menus, the  objective function of Problem (\ref{Problem3}) as a function of $v\in\cl{V}_0$ is given by
\begin{align}
J[v]&=\int_{\cl{T}}[F_{(a,k)}(\lambda^*_{(a,k)}(v),d^*_{(a,k)}(v))-v(a)]\bQ(da\times dk)\notag \\
&=\int_{\cl{T}_+ }[v'_+(a)(a-\theta^*_k - H(X_k-\theta^*_k)_+)-v(a)]\bQ(da\times dk)\notag \\
&\qquad+\int_{\cl{T}_- }[v'_-(a)(a-\theta^*_k - H(X_k-\theta^*_k)_+)-v(a)]\bQ(da\times dk)\notag \\
&=\int_{\cl{T}_+ }[v'_+(a)(a-\xi_k)-v(a)]\bQ(da\times dk)\notag \\
&\qquad+\int_{\cl{T}_- }[v'_-(a)(a-\xi_k)-v(a)]\bQ(da\times dk)
,\label{J3''}
\end{align}
in which the last equality follows from the definition of $\xi_k$.

Recall that in the previous section we introduced  $\phi_t(\cdot):=(\cdot-t)_+\in\mathcal{V}$ for any $t\in\overline{\mathbb{R}}_+$. The next result is similar to Proposition \ref{Jv} in that for each $v\in\mathcal{V}_0$ it expresses $J[v]$ in (\ref{J3''})  as a weighted average of $J[\phi_t]$.

\begin{proposition}\label{Jv:CL}
For every $v\in\mathcal{V}_0$,   $J[v]$ in (\ref{J3''}) can be written as
\[J[v]=\int_{[L,\infty)}J[\phi_t]\gamma_v(dt),\]
where $\gamma_v$ is  the Radon measure  on $[L,\infty)$ defined by $\gamma_v[L ,a]=v'_+(a)$ for $a\geq L$.
\end{proposition}
{\em Proof. } The proof is identical to that of Proposition \ref{Jv}, upon replacing $H[X_k]$, $P$, $N$, and $[0,\infty)$ there
by $\xi_k$, $\cl{T}_+$,  $\cl{T}_-$, and $[L,\infty)$ respectively. \hfill $\square$

With all these preparations,  we can now identify the optimal indirect utility for Problem (\ref{Problem3}). For every finite $t\geq L$, the corresponding objective function can be written as
\[
J[\phi_t]=\int_{\mathcal{T}}\left\{(t-\xi_k)\mathbf{1}_{\{a>t\}}+(t-\xi_k)_+\mathbf{1}_{\{a=t\}}\right\}\mathbb{Q}(da\times dk).
\]
Moreover, Assumption (\ref{OnlyAssumption}) and the monotonicity of $H$ imply that
\[
\xi_k=\theta_k^*+H[(X_k-\theta_k^*)_+]\leq L+H[X_k].
\]
Thus, by Condition (\ref{Integrability}) and the same argument as in Lemma \ref{L:QS-existence}, the map $t\mapsto J[\phi_t]$ is upper semicontinuous on $[L,\infty)\cup\{\infty\}$, with $\phi_\infty\equiv0$ and $J[\phi_\infty]=0$. Hence its maximum is attained.
\begin{theorem}\label{T:CL:v}
Define
\begin{equation}\label{CL:tau*}
\tau^*\in\operatorname*{arg\,max}_{t\in[L,\infty)\cup\{\infty\}}J[\phi_t].
\end{equation}
Then, $\max_{v\in\cl{V}_0}J[v]=J[\phi_{\tau^*}]$,
that is, the optimal indirect utility $v^*$ for Problem (\ref{Problem3}) is given by $\phi_{\tau^*}$.
\end{theorem}
{\em Proof. } Since $\tau_{\phi_{\tau^*}} =\tau^*\geq L$, $\phi_{\tau^*}\in\cl{V}_0$. Moreover, $J[\phi_{\tau^*}]\geq J[\phi_\infty]=0$. From Proposition \ref{Jv:CL},
\[J[v]=\int_{[L,\infty)}J[\phi_t]\gamma_v(dt)\leq \gamma_{v}[L,\infty)J[\phi_{\tau^*}]\leq J[\phi_{\tau^*}],\quad v\in\cl{V}_0.\]
This proves the optimality of $\phi_{\tau^*}$ within $\cl{V}_0$.
\hfill $\square$

After identifying the optimal indirect utility and with the use of Proposition \ref{CL-Op}, we can now present the optimal menu of change-loss policies for Problem (\ref{Problem3}) as an immediate consequence.

\begin{theorem}\label{T:CL} Let $\tau^*$ be defined as in (\ref{CL:tau*}).
For Problem (\ref{Problem3}), the optimal coinsurance rates  $\lambda_{(a,k)}^*$, the optimal deductibles $d^*_{(a,k)}$, and the optimal premiums $P^*_{(a,k)}$ are given respectively by
\[\lambda_{(a,k)}^* =
\begin{cases}
1, & \text{if }a>\tau^*; \\
1, & \text{if }a=\tau^*\geq\xi_k; \\
0, & \text{if }a=\tau^*<\xi_k; \\
0, & \text{if }a<\tau^*,
\end{cases}
\qquad
d_{(a,k)}^* =
\begin{cases}
\theta_k^*, & \text{if }a\geq\tau^*; \\
\text{irrelevant}, & \text{if }a<\tau^*,
\end{cases}\]
and
\[P_{(a,k)}^* =\lambda_{(a,k)}^*(a-d_{(a,k)}^*)_+- v^*(a)
=
\begin{cases}
\tau^*-\theta_k^*, & \text{if }a>\tau^*; \\
\tau^*-\theta_k^*, & \text{if }a=\tau^*\geq\xi_k; \\
0, & \text{if }a=\tau^*<\xi_k; \\
0, & \text{if }a<\tau^*.
\end{cases}\]
\end{theorem}
In fact, under Assumption (\ref{OnlyAssumption}), Theorem \ref{T:CL:v} asserts that the optimal indirect utility function must take the form of $v^*(\cdot)=\phi_t(\cdot)=(\cdot-t)_+$. For such $v^*$, it must hold that $v_+'$ and $v_-'$ are either 0 or 1. It then follows from Proposition \ref{CL-Op} that the optimal coinsurance rates are either 0 or 1 as well. Any change-loss policy with coinsurance rate being either 0 or 1 is simply a stop-loss policy. Therefore, the optima menu of  change-loss policies presented in Theorem \ref{T:CL} is the same as the   optima menu of  stop-loss policies presented in Theorem \ref{T-SL}.

The following example illustrates the result of Theorem \ref{T:CL}.
\begin{example}\label{EG:CL}
We adopt the setting and the baseline $\gamma=0.5$ from Example \ref{EG:ST}. Recall that
$\theta_k^*\approx0.170941k$ and $\xi_k\approx1.366310k$. Moreover,
\[
L=5000[-\log(0.025)]\approx18444.40
>\sup_{k\in K}\theta_k^*\approx4273.53,
\]
so Assumption (\ref{OnlyAssumption}) is satisfied. Theorem \ref{T:CL} therefore implies that the optimal change-loss threshold and indirect utility coincide with their stop-loss counterparts:
\[
\tau^*_{\rm CL}=\tau^*_{\rm SL}\approx61087.34,
\qquad
v^*(a)=(a-61087.34)_+.
\]
The optimal menu can be written compactly as
\[
(\lambda_{(a,k)}^*,d_{(a,k)}^*,P_{(a,k)}^*)=
\begin{cases}
(1,\,0.170941k,\,61087.34-0.170941k), & a\geq61087.34,\\
(0,\,\text{arbitrary},\,0), & a<61087.34.
\end{cases}
\]
Consequently, the average optimal profit, participation rate, and average information rent are respectively $16831.00$, $49.80\%$, and $10327.74$, exactly as in Example \ref{EG:ST}. The mean-matched degenerate-$\alpha$ case also carries over: the threshold is $61233.08$, and a change-loss contract with $\lambda=1$---equivalently, a stop-loss policy with deductible $0.170941k$---is offered precisely when $k\geq15015.40$.

Although the change-loss class permits partial coinsurance, the optimum occurs only at the boundary values: $\lambda=1$ for participating insurers and $\lambda=0$ for excluded insurers. Hence, under Assumption (\ref{OnlyAssumption}), enlarging the feasible contract class beyond stop-loss reinsurance does not increase the reinsurer's profit. Instead, the optimal mechanism preserves the same tradeoff between information rents and market exclusion, as well as the same $k$-dependent screening among participating types. Moreover, for every $\gamma\in[0,1]$,
\[
\sup_{k\in K}\theta_k^*\leq0.358836\times25000
=8970.91<L=18444.40,
\]
so the sensitivity results in Table \ref{tab:SL-sensitivity} carry over as well.
\begin{table}[H]
\centering
\caption{Baseline outcomes across contract classes ($\gamma=0.5$)}
\label{tab:contract-comparison}
\begin{tabular}{l|ccc}
Contract class & Average profit & Participation & Allocation \\ \hline
Stop-loss & $16831.00$ & $49.80\%$ & $k$-dependent deductible \\
Quota-share & $16744.61$ & $49.71\%$ & Common full reinsurance \\
Change-loss & $16831.00$ & $49.80\%$ & Stop-loss ($\lambda=1$)
\end{tabular}
\end{table}
Table \ref{tab:contract-comparison} summarizes the baseline comparison across the three contract classes. Although stop-loss and quota-share reinsurance generate similar aggregate participation patterns, their screening structures differ markedly: stop-loss contracts continue to differentiate participating insurers by $k$, whereas quota-share contracts pool them at a common full-reinsurance allocation. The change-loss class, despite its greater contractual flexibility, replicates the stop-loss allocation exactly and therefore yields no additional profit under Assumption (\ref{OnlyAssumption}).
\end{example}

\section{Conclusion}\label{sec:con}
This paper  solved the problem of how an reinsurer can optimally design menus of reinsurance policies under the PA framework, when
faced with agents (insurance companies) with possibly spectrum of risk characteristics and risk appetites who  adopt the V$@$R
as their risk managerial assessment in accordance with Solvency II. The reinsurer overs different
classes of reinsurance contracts, including the class
of stop-loss, quota-share, and change-loss contracts. In all three classes, the
indirect utility functions have been shown to be always in the stop-loss form. This suggests that the reinsurer will discriminate the agents  broadly into two groups according to the size of their risk exposures as measured by the levels of V$@$R prior the purchase of reinsurance. Shut-down (null) policies are offered to the low risk group, while for the high risk group, agents are either bundled, as in the class of quota-share policies, or further differentiated according to agents' risk characteristics, as in the classes of stop-loss and change-loss policies.

Several research directions can be considered as future work. First, the indirect utility plays a key role in deriving the optimal menu under the setting of continuum V$@$R-minimizers, which goes  out of sight on its usage when considering a continuum types of TV$@$R- or convex-distortion-risk-measures-minimizers. Second, it would be also of interest to check the validity of the solutions when the cost functional $H$ also varies with the types of the agents. Third, noting that the premium is treated as a decision variable in the present study, the menu might be different when restricting the premium into some well-known premium principles.


\section*{Acknowledgments}
The authors thank Dr. Ka Chun Sung for helpful discussions on some preliminary results of this work. Ka Chun Cheung was partially supported by a grant from the Research Grants Council of the Hong Kong Special Administrative Region, China (Project No.\ 17303721). Phillip Yam acknowledges the financial supports from HKGRF-14301321 with the project title ``General Theory for Infinite Dimensional Stochastic Control: Mean field and Some Classical Problems'', and HKGRF-14300123 with the project title ``Well-posedness of Some Poisson-driven Mean Field Learning Models and their Applications''. The work described in this article was also supported by a grant from the Germany/Hong Kong Joint Research Scheme sponsored by the Research Grants Council of Hong Kong and the German Academic Exchange Service of Germany (Reference No. G-CUHK411/23). He also thanks The University of Texas at Dallas for the kind invitation to be a Visiting Professor in Naveen Jindal School of Management. Yiying Zhang acknowledges the financial support from the National Natural Science Foundation of China (No. 12571513), and Shenzhen Science and Technology Program (No. JCYJ20230807093312026, No. JCYJ20250604144210013).

\section*{Declarations of interest}
No potential competing or conflict interests were reported by the authors.

\setlength{\bibsep}{0.0pt}
\bibliography{adverse_selection_continuum}

\end{document}